\pdfoutput=1 

\documentclass[reprint,superscriptaddress,nofootinbib,amsmath,amssymb,aps,prl,floatfix]{revtex4-2}
\usepackage{graphicx}
\usepackage{dcolumn}
\usepackage{bm}
\usepackage{graphicx} 
\usepackage{xcolor}
\usepackage{svg} 
\usepackage{siunitx} 
\usepackage{booktabs} 
\usepackage{amsmath}

\usepackage{hyperref}
\usepackage{cleveref} 

\usepackage{orcidlink} 

\usepackage[nolist]{acronym} 

\newcommand{\rev}[1]{{\color{black} #1}}

\begin{document}

\preprint{APS/123-QED}

\title{Squeezing at the normal-mode splitting frequency of a nonlinear coupled cavity}

\author{Jonas Junker \orcidlink{0000-0002-3051-4374}}
\email[Corresponding author: ]{jonju@dtu.dk}
\affiliation{OzGrav, Centre for Gravitational Astrophysics, Research School of Physics \& Research School of Astronomy and Astrophysics, Australian National University, Australian Capital Territory, Australia.}
\affiliation{Center for Macroscopic Quantum States (bigQ), Department of Physics, Technical University of Denmark, 2800 Kongens Lyngby, Denmark}

\author{Jiayi Qin \orcidlink{0000-0002-7120-9026}}
\affiliation{OzGrav, Centre for Gravitational Astrophysics, Research School of Physics \& Research School of Astronomy and Astrophysics, Australian National University, Australian Capital Territory, Australia.}

\author{Vaishali B. Adya \orcidlink{0000-0003-4955-6280}}
\affiliation{Nonlinear and Quantum Photonics Lab, Department of Applied Physics, KTH Royal Institute of Technology, Stockholm, Sweden.} 

\author{Nutsinee Kijbunchoo \orcidlink{0000-0002-2874-1228}}
\affiliation{OzGrav, University of Adelaide, Adelaide, South Australia, Australia.}

\author{Sheon S. Y. Chua \orcidlink{0000-0001-8026-7597}}
\affiliation{OzGrav, Centre for Gravitational Astrophysics, Research School of Physics \& Research School of Astronomy and Astrophysics, Australian National University, Australian Capital Territory, Australia.}

\author{Terry G. McRae \orcidlink{0000-0002-6540-6824}}
\affiliation{OzGrav, Centre for Gravitational Astrophysics, Research School of Physics \& Research School of Astronomy and Astrophysics, Australian National University, Australian Capital Territory, Australia.}

\author{Bram J. J. Slagmolen \orcidlink{0000-0002-2471-3828}}
\affiliation{OzGrav, Centre for Gravitational Astrophysics, Research School of Physics \& Research School of Astronomy and Astrophysics, Australian National University, Australian Capital Territory, Australia.}

\author{David E. McClelland \orcidlink{0000-0001-6210-5842}}
\affiliation{OzGrav, Centre for Gravitational Astrophysics, Research School of Physics \& Research School of Astronomy and Astrophysics, Australian National University, Australian Capital Territory, Australia.}


\date{\today}

\begin{abstract}
Coupled optical cavities, which support normal modes, play a critical role in optical filtering, sensing, slow-light generation, and quantum state manipulation. Recent theoretical work has proposed incorporating nonlinear materials into these systems to enable novel quantum technologies. Here, we report the first experimental demonstration of squeezing generated in a quantum-enhanced coupled-cavity system, achieving a quantum noise reduction of \rev{\SI{3.3}{\decibel} around the} normal-mode splitting frequency of \SI{7.47}{\mega\hertz}. We provide a comprehensive analysis of the system’s loss mechanisms and performance limitations, validating theoretical predictions. Our results underscore the promise of coupled-cavity squeezers for advanced quantum applications, including gravitational wave detection and precision sensing.

\end{abstract}

\maketitle

\begin{acronym}
    \acro{OPO}{optical parametric oscillator}
    \acro{PPKTP}{periodically-poled potassium titanyl phosphate}
    \acro{FSR}{free spectral range}
    \acro{SQL}{standard quantum limit}
    \acro{PZT}{piezoelectric transducer}
    \acro{GWD}{gravitational wave detector}
    \acro{LO}{local oscillator}
    \acro{HD}{homodyne detector}
    \acro{HR}{high-reflective}
    \acro{FP}{Fabry–Pérot}
    \acro{SHG}{second harmonic generation}
\end{acronym}

Coupled oscillators exhibit unique properties, such as the resonances of normal modes observable in simple mechanical systems like two masses connected by a spring \cite{saulson1990ThermalNoiseMechanical}. Normal modes also arise in more complex systems, such as optomechanical oscillators \cite{dobrindt2008ParametricNormalModeSplitting}, 
light-atom interactions \cite{thompson1992ObservationNormalmodeSplitting,friskkockum2019UltrastrongCouplingLight}, and coupled cavities. These coupled oscillators exhibit a rich spectrum of interesting phenomena. For example, in quantum optomechanics, cavities with membranes or cantilevers \cite{aspelmeyer2014CavityOptomechanics}, enhance displacement sensitivity \cite{dobrindt2010TheoreticalAnalysisMechanical}, enable oscillator cooling \cite{chan2011LaserCoolingNanomechanical}, and exploit quantum back-action \cite{moller2017QuantumBackactionevadingMeasurement}. Atom-light interactions provide platforms for quantum interfaces \cite{hammerer2010QuantumInterfaceLight} or evade quantum back-action \cite{moller2017QuantumBackactionevadingMeasurement}.

Normal-mode splitting in coupled cavities arises from the constructive and destructive interference of optical modes. This phenomenon enables applications such as narrow-band filtering and precise mode selection \cite{lu2012TunableHighchannelcountBandpass}, coupled-cavity-induced transparency \cite{smith2004CoupledresonatorinducedTransparency,zheng2012ControllableOpticalAnaloga,guo2021TransitionCoupledresonatorinducedTransparencya}, sensing \cite{righini2016BiosensingWGMMicrospherical}, slow light generation \cite{totsuka2007SlowLightCoupledResonatorInduced}, and manipulation of squeezed states \cite{di2011CoupledResonatorInducedTransparencySqueezed}. The mode-splitting parameters depend solely on the optical properties of the cavities \cite{li2010CoupledModeTheory}, offering high tunability, robustness for on-chip photonic solutions, and operation at room temperature.

The integration of nonlinear materials into coupled optical cavities has emerged as a promising area of research, inspiring various theoretical proposals. A key application is internal squeezing \cite{korobko2023FundamentalSensitivityLimit,gardner2022NondegenerateInternalSqueezinga,somiya2016ParametricSignalAmplificationa,miao2019QuantumLimitLaser,rehbein2005OpticalTransferFunctions,peano2015IntracavitySqueezingCan,korobko2017BeatingStandardSensitivityBandwidth,korobko2023MitigatingQuantumDecoherence,bottner2024CoherentFeedbackQuantum}, where a nonlinear crystal is placed within the coupled cavity of a \ac{GWD}, enhancing sensitivity to events such as binary neutron star mergers detectable at the system's normal-mode frequency \cite{korobko2019QuantumExpanderGravitationalwave,adya2020QuantumEnhancedKHz}. Other studies explore enhancing second-harmonic generation by incorporating second-order nonlinear materials into one cavity of a coupled microcavity system \cite{lv2022GainEnhancedSecond}. Effects such as unconventional amplification \cite{huang2017DualismOpticalDifference} or photon blockades \cite{liu2020PhotonBlockadeEnhancing} seem to be also feasible when a nonlinear material is placed in a coupled cavity system. Most recent work suggests that squeezing can be amplified by coupling two optical cavities containing nonlinear materials \cite{jabri2024LightSqueezingEnhancement}. However, realizing these concepts experimentally remains a critical challenge for unlocking their full potential.

We present a system consisting of a squeezing cavity coupled to an empty test cavity, measuring \rev{\SI{3.3}{\decibel}} of squeezing \rev{around} the system's normal-mode splitting frequency of \SI{7.47}{\mega\hertz}. This quantum-enhanced non-linear coupled cavity is locked by two nominally decoupled modes. Through a thorough characterization of all loss channels, we identify the setup's challenges and limitations, such as the system's escape efficiency and phase noise. Our measurements validate the theoretical predictions from \cite{korobko2019QuantumExpanderGravitationalwave}, demonstrating that a coupled cavity squeezer generates squeezing at the normal-mode frequencies. 
This result marks an important milestone in advancing nonlinear quantum-enhanced coupled cavities, especially to enable next-generation high-frequency gravitational wave detectors. Our system could be also interesting for quantum information processing or quantum memory.

\begin{figure}[htbp]
  \centering
  \includegraphics[width=1\columnwidth]{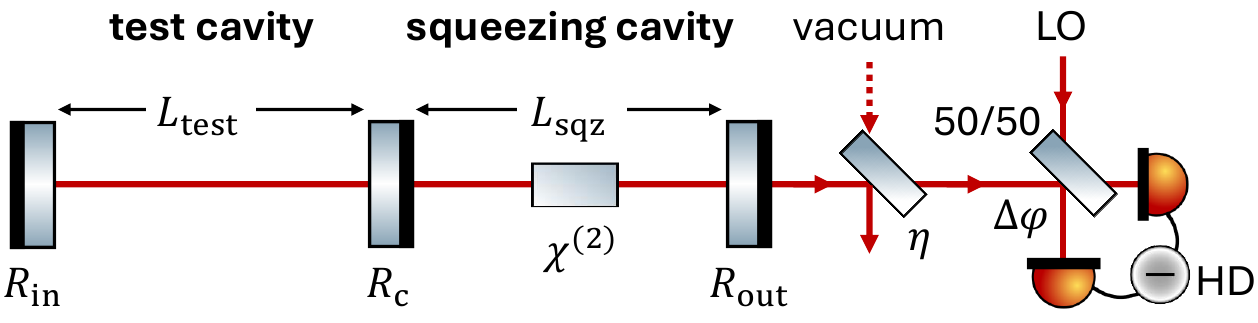}
  \caption{Conceptual diagram of a test cavity coupled to a squeezing cavity. The squeezed output state undergoes losses of \( 1-\eta \) and is directed to a balanced homodyne detector (HD), where it interferes with a local oscillator (LO) and phase noise \( \Delta \varphi \) is present.}
  \label{fig:theoretical_setup}
\end{figure}

\Cref{fig:theoretical_setup} illustrates a conceptual coupled-cavity system consisting of a test cavity coupled to a squeezing cavity via a mirror with power reflectivity \( R_\text{c} \). The squeezing cavity incorporates a nonlinear \( \chi^{(2)} \) medium with down-conversion coupling strength \(\chi \propto \chi^{(2)}\). The optical lengths of the test and squeezing cavities are \( L_\text{test} \) and \( L_\text{sqz} \), respectively. The system's input and output power reflectivities are \( R_\text{in} \) and \( R_\text{out} \), respectively, and its decay rate is denoted by \(\gamma\). When the test cavity is blocked, the system reduces to the squeezing cavity alone with decay rate $\gamma_\text{sqz}$. Then, the output noise variances of the amplitude (\(+\)) and phase (\(-\)) quadratures, dependent on the measurement frequency $\omega$ are \cite{walls2008QuantumOptics,collett1984SqueezingIntracavityTravelingwavea,collett1985SqueezingSpectraNonlinear}
\begin{equation}
\label{eq:HFsqz}
    V_\text{sqz}^\pm (\omega) = 1 \pm \frac{4 \gamma_\text{sqz} \chi}{(\gamma_\text{sqz} \mp \chi)^2 + \omega^2}. 
\end{equation}

For the coupled system, the amplitude noise transfer function to the output can be approximated as \cite{korobko2019QuantumExpanderGravitationalwave}
\begin{equation}
    \mathcal{T}^\pm(\omega) = \frac{(\gamma \mp \chi) \omega + i (\omega^2-\omega_\text{s}^2)}{(\gamma \pm \chi) \omega + i (\omega^2-\omega_\text{s}^2)},
\end{equation}
resulting in output noise variances given by
\begin{align}
    V_\text{c}^\pm (\omega) &=  |\mathcal{T}^\pm(\omega)|^2 \\&= 1 \pm \frac{4 \gamma \chi \omega^2}{(\gamma\mp\chi)^2\omega^2+\omega^4-2\omega^2 \omega_\text{s}^2+\omega_\text{s}^4}.
\end{align}

The noise variances reveal a quantum noise reduction at \( \omega_\text{s} \), the normal-mode splitting frequency resulting from the coupling between the two cavities. In the approximation of small couplings \( T_\text{c} = 1-R_\text{c} \ll 1 \), this frequency depends on the optical lengths of the individual cavities and the transmissivity of the coupling mirror \cite{devine2003MeasurementFrequencyResponse}:
\begin{equation}
    \omega_\text{s} = c \sqrt{\frac{T_\text{c}}{4 L_\text{test} L_\text{sqz}}}.
\end{equation}

Upon exiting the coupled cavity, the quantum state experiences total optical losses of \( 1-\eta \), which are described by a beam splitter process. At the beam splitter of the \ac{HD}, the detection angle \( \Delta \varphi \) introduces additional phase noise. These two effects degrade the measured variance of the readout quadrature, with $i \in \{\text{sqz}, \text{c} \}$: 
\begin{equation}
\label{eq:fitting_model}
    V^\pm_\text{det} = \left( V_i^\pm \cos^2 \Delta \varphi + V_i^\mp \sin^2 \Delta \varphi \right) \eta + 1 - \eta.
\end{equation}
\Cref{eq:fitting_model}, which accounts for phase noise, optical losses, and frequency dynamics, is used to fit the measured squeezed noise variances shown later.

\begin{figure}[htbp]
  \centering
  \includegraphics[width=1\columnwidth]{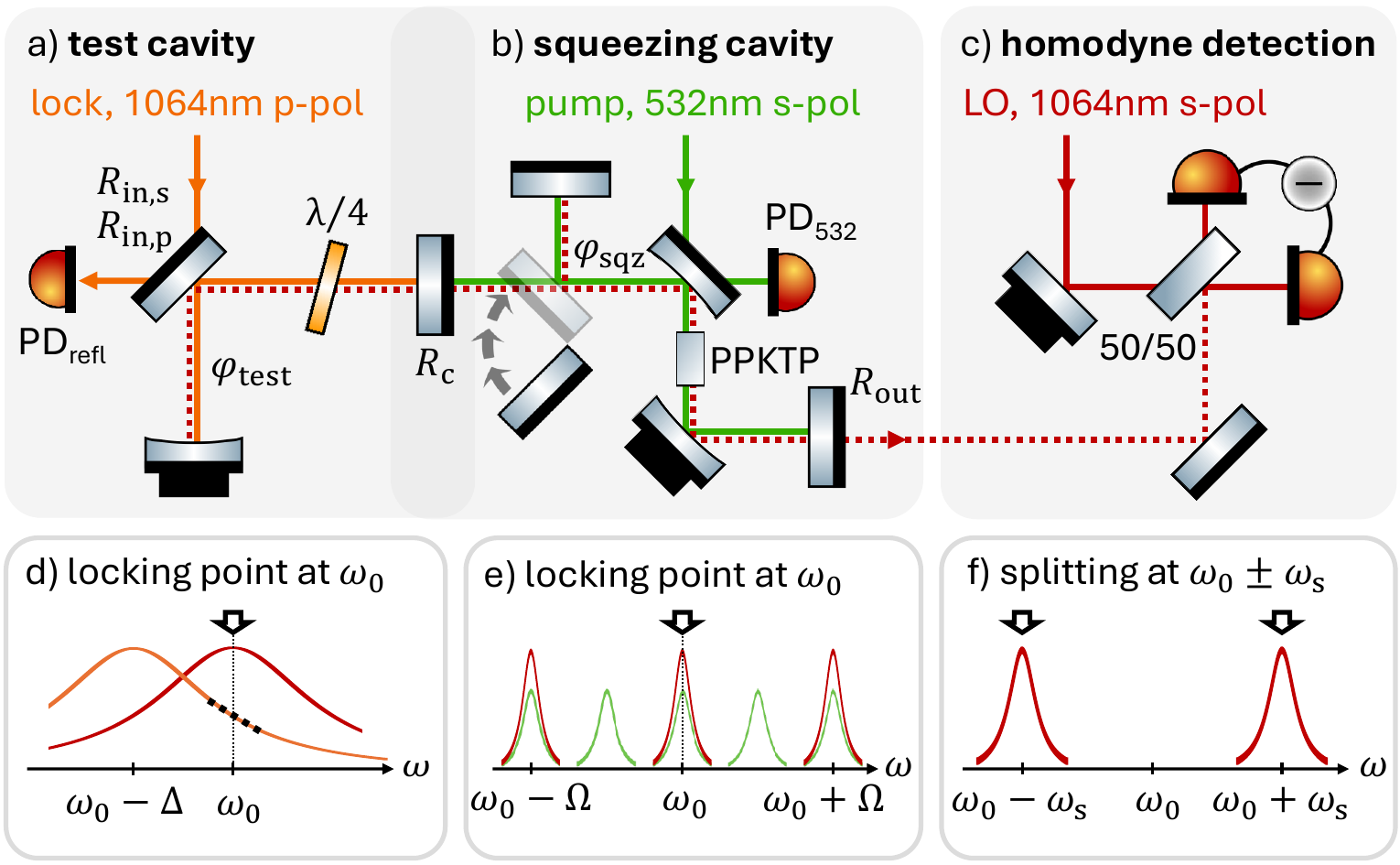}
  \caption{Top row: Experimental setup diagram including a) the test cavity, b) the squeezing cavity and c) the homodyne detection. When the flipping mirror is up, the squeezing cavity is decoupled from the test cavity. Bottom row: Resonance responses of the three contributing modes in d) the test cavity with blue dotted error signal, e) the squeezing cavity independently, and f) the coupled oscillator’s resonant modes as detected by the homodyne detector.}
  \label{fig:experimental_setup}
\end{figure}
Our experimental setup, shown in \Cref{fig:experimental_setup}, comprises three main subsystems.

The first subsystem is a folded \ac{FP} test cavity with an effective length of \( L_\text{test} = \SI{1.415}{\meter} \) that is stabilized using a p-polarized field. To modify the phase degeneracy between the two polarization modes, a slightly tilted quarter-wave plate is incorporated into the test cavity. The waveplate is aligned with its fast axis parallel to the cavity’s polarization basis, ensuring no coupling is introduced between the polarizations. A p-polarized light field at the fundamental wavelength of \( \lambda = \SI{1064}{\nano\meter} \) enters the test cavity through the input mirror, which has polarization-dependent power reflectivities of \( R_\text{in,s} = \SI{99.6}{\percent} \) and \( R_\text{in,p} = \SI{99.0}{\percent} \). The test cavity includes \rev{a} \ac{HR} mirror that is actuated by a \rev{\ac{PZT}} to control the cavity phase $\phi_\text{test}$. This curved mirror is the only focusing element, setting the cavity eigenmode waist at the coupling mirror with reflectivity \( R_\text{c} = \SI{81.0}{\percent} \). The coupling mirror serves as the interface between the test cavity and the second subsystem, the squeezing cavity.

The squeezing cavity is a folded \ac{FP} cavity with an optical path length of \( L_\text{sqz} = \SI{1.372}{\meter} \), \rev{resulting in a \ac{FSR} of $2 \pi \times \Omega = \SI{109.34}{\mega\hertz}$.} This doubly-resonant cavity supports circulating fields for the s-polarised fundamental field and the pump at \SI{532}{\nano \meter}, generated by an external \ac{SHG} cavity (not shown in \Cref{fig:experimental_setup}). Two \ac{HR} curved mirrors with radii of curvature \( \text{ROI} = \SI{100}{\milli \meter} \) are used to focus the beam into a \( \SI{1}{\milli \meter} \times \SI{5}{\milli \meter} \times \SI{10}{\milli \meter} \) \ac{PPKTP} crystal. These mirrors are hit by the beam under an angle of incidence of roughly \SI{8}{\degree} (not represented in \Cref{fig:experimental_setup}). The \ac{PPKTP} crystal is wedged along its \SI{5}{\milli \meter} horizontal side by \( \SI{1.43}{\degree} \) and is temperature-controlled at \( \SI{30}{\celsius} \) to optimize phase-matching for efficient nonlinear down-conversion. The cavity ends with an output mirror of reflectivity \( R_\text{out} = \SI{90.3}{\percent} \), which couples the squeezed light out of the cavity.

An \ac{HR} mirror can be flipped up to decouple the cavities and independently characterize the squeezer. This configuration preserves the squeezer's cavity length and spatial mode parameters, ensuring consistency in the optical characteristics. Characterizing the squeezing cavity by blocking the test cavity would yield an escape efficiency of \( \eta_\text{esc} < \SI{34}{\percent} \), making this approach unsuitable for accurate characterization. 

The final subsystem is the balanced \ac{HD}, which detects the squeezed state. The photodetector is an in-house design with a bandwidth of approximately \SI{100}{\mega\hertz}, directly subtracting the photocurrents of the two photodiodes. The \ac{LO} phase is controlled using a \ac{PZT}-actuated mirror, and the \ac{LO} optical power is maintained at around \SI{12}{\milli \watt} throughout the measurements. At this power, the homodyne detector still works linear and has a high shot noise clearance at relevant frequencies.

To generate a squeezed state at the normal-mode splitting frequency, stable locking of the entire system is essential. Normal-mode splitting occurs when both cavities are independently resonant for the fundamental frequency. The strong coupling (\( T_\text{c} = \SI{19}{\percent} \)) complicates the separation of individual cavity responses. While recent methods propose using phase modulation and beat-frequency demodulation to decouple error signals \cite{maggiore2024TuningResonantDoublets}, our approach leverages a locking technique based on two distinct modes with different wavelengths and polarizations. This method successfully decouples the systems and generates independent error signals, enabling stable locking for squeezed state generation.

The test cavity phase \( \varphi_\text{test} \) is locked using a p-polarized field at the fundamental frequency, as depicted in \Cref{fig:experimental_setup}d. To set the relative phase between the p- and s-polarized cavity modes \( \Delta \), we adjust the horizontal tilt of the quarter-wave plate mounted on a motorized rotation stage [Thorlabs, ELL18]. This tilt modifies the differential optical path length, thereby controlling the relative phase between the polarizations. The side-fringe slope of the cavity's p-polarized amplitude response, measured on photodetector PD$_\text{refl}$, serves as an error signal to maintain resonance for the s-polarized mode. The low finesse of the squeezing cavity for the fundamental p-polarization ensures negligible impact of the squeezer's cavity dynamics on the phase control of the test cavity. Thus the p-polarisation does not see a coupled cavity. 

The squeezing cavity phase \( \varphi_\text{sqz} \) is locked with the pump field using a modulation-free polarization-based homodyne technique (indicated by PD$_{532}$ in \Cref{fig:experimental_setup}) \cite{hansch1980LaserFrequencyStabilization}. This lock is independent of the 
test cavity dynamics due to the low intra-cavity mirror reflectivities and the \ac{HR} coupling mirror for the pump. The same locking method is used when the coupling mirror is replaced by the path through the flipped-up HR mirror. The fundamental s-polarized field can be tuned to resonance by displacing the wedged birefringent crystal, adjusting the optical path length and relative phase with the pump field. Fine-tuning is achieved with small temperature changes to the crystal, which minimally affect the down-conversion phase matching.

Our locking routine allows arbitrary control of the fundamental resonance condition. For instance, the fundamental can be co-resonant (squeezing at DC and \( \Omega \)), as shown in \Cref{fig:experimental_setup}e, or anti-resonant (squeezing at \( \Omega/2 \)). \rev{In this plot, the green peaks are only projections for better visualization of the co-resonaces.} With both cavities independently locked on resonance, the coupled system enhances and correlates modes at the splitting frequency \( \pm \omega_\text{s} \), as illustrated in \Cref{fig:experimental_setup}f.

To independently test and characterize the squeezer, we first performed a squeezing measurement by flipping the \ac{HR} mirror up into the beam path. This replaced the coupling mirror, isolating the squeezer from the test cavity. For this measurement, we used \SI{55}{\milli \watt} of pump power to see a fair amount of squeezing but also be sensitive to phase noise. We locked the squeezer to the pump field and set the relative phase to achieve anti-resonance for the fundamental mode. As a result, squeezing was generated at half the \ac{FSR}, \( \Omega/2 \). At this frequency, we are neither limited by low-frequency technical noise nor by the limited bandwidth of the \ac{HD}.

\begin{figure}[htbp]
	  \centering
        \includegraphics[width=1\columnwidth]{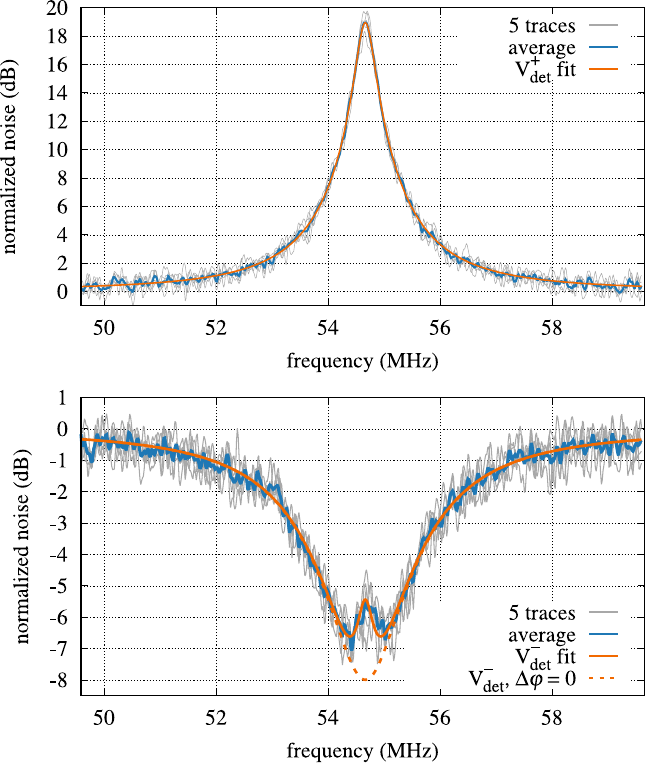}
\caption{Anti-squeezing (top) and squeezing (bottom) at half the \ac{FSR} ${\Omega/2 = 2\pi \times \SI{54.67}{\mega \hertz}}$, normalized to shot noise. The five measured spectra with highest and lowest noise are shown in grey. We averaged (\rev{blue} traces) and fitted the data using \Cref{eq:fitting_model} (\rev{orange} traces, dashed trace assumes no phase noise). At the center frequency, anti-squeezing couples into the squeezed readout quadrature because of phase noise resulting in the local maximum. The electronic dark noise, which was \SI{12.7}{\deci \bel} below the shot noise, was subtracted from the noise data. Used pump power: $P = \SI{55}{\milli \watt}$. The fitting parameters are: $\Delta \varphi = \SI{40}{\milli \radian}$ and $\eta = \SI{85.1}{\percent}$, $\Omega = 2 \pi \times \SI{109.34}{\mega \hertz}$, $\chi =2 \pi \times \SI{704}{\kilo \hertz}$, $\gamma_\text{sqz} = 2 \pi \times \SI{868}{\kilo \hertz}$.}
\label{fig:meas_55MHz}
\end{figure}

Because of free-running pump and detection phases, we measured 1000 spectra over a span of \SI{10}{\mega \hertz} at half the \ac{FSR} frequency of \SI{54.67}{\mega \hertz}. To analyze the data, we selected the five spectra with the highest and lowest noise levels (grey traces) and averaged them separately (\rev{blue} traces), as shown in \Cref{fig:meas_55MHz}. We subtracted the photodetector's dark noise from the measurements and normalized the data to shot noise. To characterize the data, we fitted the model given by \Cref{eq:HFsqz} inserted into \Cref{eq:fitting_model}, with \(\omega_0 = \Omega/2\) to both averaged traces simultaneously (\rev{orange} traces). The fit results reveal a maximum anti-squeezing of \SI{18.9}{\deci\bel} above shot noise, and a minimum squeezing of \SI{6.6}{\deci\bel} below shot noise. Additionally, it yields values for the total efficiency $\eta = \SI{85.1}{\percent}$ and phase noise $\Delta \varphi = \SI{40}{\milli \radian}$. The observed phase noise is likely due to the $\SI{1.372}{\meter}$-long free-space squeezer, which contains ten intra-cavity components and lacks phase-locking mechanisms. For comparison, we also plot the model function for $V^-_\text{det}$ without phase noise ($\Delta \rev{\varphi} = 0 $, \rev{dashed} orange trace). From the fitted decay rate $\gamma_\text{sqz} = 2 \pi \times \SI{868}{\kilo \hertz}$ and nonlinear efficiency $\chi = 2 \pi \times \SI{704}{\kilo \hertz}$, we calculated the squeezer's oscillation threshold as $P_\text{thresh} = P \,  \gamma_\text{sqz}^2/\chi^2 \approx \SI{83.7}{\milli \watt}$.

With the other loss channels measured independently, we performed a thorough characterization of the squeezing setup. In separate measurements, we determined an escape efficiency of \(\eta_\text{esc} = \SI{97.5}{\percent}\), a propagation efficiency of \(\eta_\text{prop} = \SI{94.0}{\percent}\), and a homodyne efficiency of \(\eta_\text{hd} = \SI{94.5}{\percent}\), as summarized in \Cref{tab:efficiencies}. Combining these efficiencies with the total efficiency from the fit, we derived a remaining loss, which we attribute to the quantum efficiency of the \ac{HD} as in \cite{vahlbruch2016Detection15DB}, with \(\eta_\text{qe} = \SI{98.2}{\percent}\). This value is consistent with the detector's gain measurements, which convert optical power into voltage.
\begin{table}
\begin{ruledtabular}
	\begin{tabular}{@{}ccc@{}}
        parameter & single squeezer & coupled system \\
        \midrule
		\midrule
  	 $\eta_\text{esc}$ ($\SI{}{\percent}$) & 97.5 & 76.8 \\
          $\eta_\text{prop}$ ($\SI{}{\percent}$) & 94.0 & 94.0 \\
          $\eta_\text{hd}$ ($\SI{}{\percent}$) & 94.5 & 89.1$^{*}$ \\
         $\eta_\text{\rev{qe}}$ ($\SI{}{\percent}$) & 98.2$^{*}$ & 98.2 \\
        \midrule
        \midrule
	    $\eta$ ($\SI{}{\percent}$) & 85.1 & 63.2 \\
 	$\Delta \varphi$ ($\SI{}{\milli \radian}$) & 40 & 79 \\
        \midrule
        \midrule
        anti-squeezing (dB) & 18.9 & 14.8 \\ 
        squeezing (dB) & $-6.6$ & $-3.3$ \\ 
	\end{tabular}
\end{ruledtabular}
\caption{Overview of the efficiencies in the two setups. The total efficiency $\eta$ and the phase noise $\Delta \phi$ are obtained from fitting the data from \Cref{fig:meas_55MHz} and \Cref{fig:meas_7.5MHz}. Values with asterisk are inferred from the overall loss budget.}
\label{tab:efficiencies}
\end{table}

\begin{figure}[htbp]
	  \centering
        \includegraphics[width=1\columnwidth]{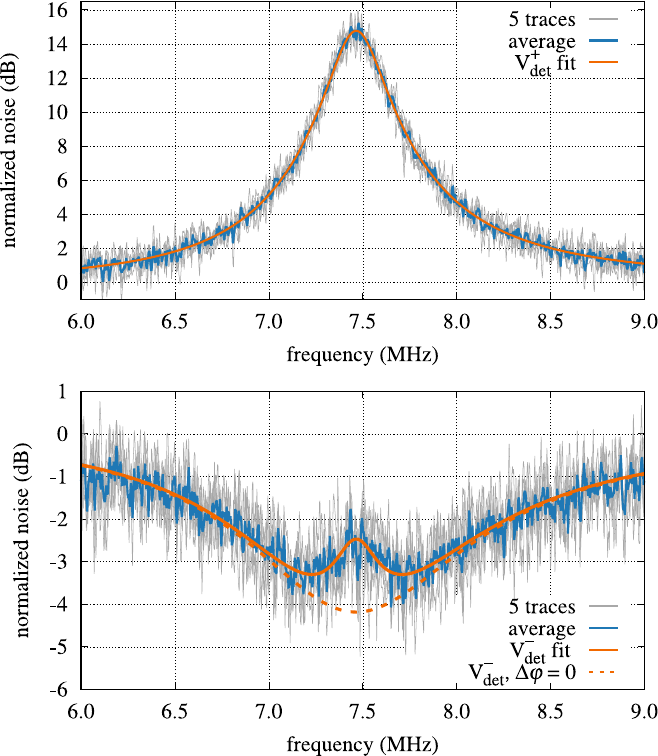}
\caption{Anti-squeezing (top) and squeezing (bottom) \rev{around} the normal-mode splitting frequency $\omega_\text{s} = 2\pi \times \SI{7.47}{\mega \hertz}$, normalized to shot noise. Measurements were taken as in \Cref{fig:meas_55MHz}. The electronic dark noise was \SI{13.2}{\deci \bel} below the shot noise and subtracted from the noise data. Used pump power: $P = \SI{74}{\milli \watt}$. Fitting parameters: $\Delta \varphi = \SI{79}{\milli \radian}$ and $\eta = \SI{63.2}{\percent}$, $\omega_\text{s} = 2 \pi \times \SI{7.47}{\mega \hertz}$, $\chi = 2 \pi \times \SI{820}{\kilo \hertz}$, $\gamma = 2 \pi \times \SI{1.100}{\mega \hertz}$.}
\label{fig:meas_7.5MHz}
\end{figure}
To study the coupled system, we flipped the \ac{HR} mirror down again and locked both cavities as described above. In contrast to the previous measurement, we locked the squeezer to be co-resonant with the pump field. We recorded 1000 noise spectra at the splitting frequency of \SI{7.47}{\mega \hertz} with a span of \SI{3}{\mega \hertz} and selected the five highest/lowest noise data sets, plotted by the grey traces \rev{in \Cref{fig:meas_7.5MHz}}. After subtracting dark noise and normalizing to shot noise, we again fitted the average of the five traces presented by the \rev{blue} traces with the highest and lowest noise with the model in \Cref{eq:fitting_model}. The fits reveal a maximum anti-squeezing of \SI{14.8}{\decibel} above the shot noise and a minimum squeezing of \SI{3.3}{\decibel} below the shot noise (\rev{orange} traces). Additionally, the fit provided values for the total efficiency \(\eta = \SI{63.2}{\percent}\) and phase noise \(\Delta \varphi = \SI{79}{\milli \radian}\). For this measurement, we used a slightly higher pump power of $P=\SI{74}{\milli \watt}$ to increase the sensitivity to phase noise, enabling a more precise characterization. The phase noise during detection increased compared to the previous setup due to the coupled system incorporating more free-space optical components, which heightened its susceptibility to intra-cavity phase noise. For future experiments, improving the table's housing and implementing phase controls for both the pump and detection phases would be highly beneficial.

The noise data from the single-cavity squeezer were fitted with a nonlinear conversion efficiency of \(\chi = 2 \pi \times \SI{704}{\kilo \hertz}\), while the coupled-cavity squeezer data were fitted with \(\chi = 2 \pi \times \SI{820}{\kilo \hertz}\). This difference aligns with the different pump powers (\(\SI{55}{\milli \watt}\), \(\SI{74}{\milli \watt}\)), which scale with the nonlinear efficiency as \(\chi \propto \sqrt{P}\). Consequently, the threshold power increased to \(P_\text{thresh} \approx \SI{133.1}{\milli \watt}\).

The overall loss budget increased due to additional losses in the coupled system and a modified homodyne contrast. The coupled setup includes new optical components, such as the quarter waveplate with losses of \(L_\text{qwp} = 2 \times \SI{0.8}{\percent}\), the test cavity's input-output mirror with \(L_\text{in,s} = 2 \times \SI{0.4}{\percent}\). The remaining losses account for \(L_\text{res} = \SI{0.3}{\percent}\). As a result, the escape efficiency reduced to \(\eta_\text{esc} = \SI{76.8}{\percent}\), consistent with the observed increase in the fitted decay rate to \(\gamma = 2 \pi \times \SI{1.100}{\mega \hertz}\). Improving the escape efficiency would require some targeted optimizations. Employing higher-quality optical components to reduce intra-cavity losses could make an escape efficiency exceeding \SI{90}{\percent} achievable. Additionally, adopting an alternative locking scheme for the test cavity to eliminate the intra-cavity quarter waveplate could further enhance escape efficiency.

Accurately measuring the homodyne contrast is challenging due to the anti-resonance of the coupled-cavity system for the carrier field, which is a consequence of the normal mode splitting. To address this, we optimized the contrast by interfering the local oscillator with a bright field reflected from the mode-matched squeezing cavity while blocking the test cavity. However, this bright field does not perfectly match the spatial mode of the coupled-cavity field, resulting in only an approximate estimation of the homodyne contrast. For a more precise contrast, we inferred the degraded homodyne efficiency from the total loss budget, yielding \(\eta_\text{hd} = \SI{89.1}{\percent}\). The contrast could be improved by better matching the eigenmodes of the two cavities, ensuring that the bright field leaking from the squeezing cavity serves as a suitable reference for the squeezed field. Alternatively, a sideband at the resonant splitting frequency could be used to generate a beat note with the local oscillator at the \ac{HD}.

In conclusion, we demonstrated \rev{\SI{3.3}{\decibel}} of quantum noise reduction generated by a coupled-cavity squeezer. We controlled the lengths of each cavity independently by separate polarisation and frequency modes, to generate the squeezing at the system's normal-mode splitting frequency of \SI{7.47}{\mega \hertz}. We presented a detailed analysis of the loss couplings and highlighted the associated challenges. For the characterization, we fitted our experimental data to the theory of a coupled-cavity squeezer \cite{korobko2019QuantumExpanderGravitationalwave}, incorporating losses and phase noise in the readout. The generated squeezing is primarily limited by the system's relatively low escape efficiency of \SI{76.8}{\percent}. However, with optimizations to the cavity design and improved optical components, an escape efficiency of at least \SI{90}{\percent} should be achievable. \rev{In addition, the propagation and homodyne detection efficiencies could realistically be increased to \SI{99}{\percent} and \SI{95}{\percent}, respectively. If then the phase noise can be stabilized below \SI{20}{\milli\radian} using phase locks, the system could achieve a maximum squeezing level of \SI{6.9}{\decibel}.} 

Our results highlight the feasibility of quantum-enhanced nonlinear coupled systems and are particularly important for informing future \ac{GWD} designs employing the concept of internal squeezing. With the addressed optimisations, demonstrating the positive effect on the signal-to-noise-ratio for internal squeezing will be possible in this coupled system. Our system could furthermore provide valuable insights into loss couplings in coupled cavities, such as those in current \acp{GWD} LIGO \cite{collaboration2015AdvancedLIGO,ligoo4detectorcollaboration2023BroadbandQuantumEnhancement},  Virgo \cite{acernese2014AdvancedVirgoSecondgeneration} and KAGRA  \cite{akutsu2021OverviewKAGRADetector,thekagracollaboration2013InterferometerDesignKAGRA}. Our insights are interesting for advanced quantum techniques related to second-harmonic generation, amplification, photon blockades and squeezing improvement. In our experiment, the sub-cavities were locked on resonance, resulting in symmetric mode splitting. We observed that detuned sub-cavities alter the generated output state, leading to frequency-dependent squeezing, similar as in \cite{junker2022FrequencyDependentSqueezingDetuned}, which warrants further exploration.

\begin{acknowledgments}
We thank Mikhail Korobko for insightful discussions and helpful remarks. We thank Daniel Gould, Dennis Wilken and Nived Johny for useful discussions and we thank Michèle Heurs from the Leibniz University Hannover for generously lending some optical components for our experiment. This research was supported by the Australian Research Council under the Discovery Grant scheme, Grant No. DP220102755. We would like to also thank the support of the Australian Research Council under the ARC Centre of Excellence for Gravitational Wave Discovery, Grants No. CE170100004, and No. CE230100016. V. B. A. would like to acknowledge the support and funding from the Swedish Research Council (VR starting Grant No. 2023-0519 and Optical Quantum Sensing environment Grant No. 201606122) and the Wallenberg Center for Quantum Technology (WACQT) in Sweden. 
\end{acknowledgments}

\section*{Data availability}
The data that support the plots within this paper and other findings of this study are available from the corresponding author upon reasonable request.

\section*{Competing Interests}
The authors declare no competing interests.


\begin{thebibliography}{45}%
\makeatletter
\providecommand \@ifxundefined [1]{%
 \@ifx{#1\undefined}
}%
\providecommand \@ifnum [1]{%
 \ifnum #1\expandafter \@firstoftwo
 \else \expandafter \@secondoftwo
 \fi
}%
\providecommand \@ifx [1]{%
 \ifx #1\expandafter \@firstoftwo
 \else \expandafter \@secondoftwo
 \fi
}%
\providecommand \natexlab [1]{#1}%
\providecommand \enquote  [1]{``#1''}%
\providecommand \bibnamefont  [1]{#1}%
\providecommand \bibfnamefont [1]{#1}%
\providecommand \citenamefont [1]{#1}%
\providecommand \href@noop [0]{\@secondoftwo}%
\providecommand \href [0]{\begingroup \@sanitize@url \@href}%
\providecommand \@href[1]{\@@startlink{#1}\@@href}%
\providecommand \@@href[1]{\endgroup#1\@@endlink}%
\providecommand \@sanitize@url [0]{\catcode `\\12\catcode `\$12\catcode `\&12\catcode `\#12\catcode `\^12\catcode `\_12\catcode `\%12\relax}%
\providecommand \@@startlink[1]{}%
\providecommand \@@endlink[0]{}%
\providecommand \url  [0]{\begingroup\@sanitize@url \@url }%
\providecommand \@url [1]{\endgroup\@href {#1}{\urlprefix }}%
\providecommand \urlprefix  [0]{URL }%
\providecommand \Eprint [0]{\href }%
\providecommand \doibase [0]{https://doi.org/}%
\providecommand \selectlanguage [0]{\@gobble}%
\providecommand \bibinfo  [0]{\@secondoftwo}%
\providecommand \bibfield  [0]{\@secondoftwo}%
\providecommand \translation [1]{[#1]}%
\providecommand \BibitemOpen [0]{}%
\providecommand \bibitemStop [0]{}%
\providecommand \bibitemNoStop [0]{.\EOS\space}%
\providecommand \EOS [0]{\spacefactor3000\relax}%
\providecommand \BibitemShut  [1]{\csname bibitem#1\endcsname}%
\let\auto@bib@innerbib\@empty
\bibitem [{\citenamefont {Saulson}(1990)}]{saulson1990ThermalNoiseMechanical}%
  \BibitemOpen
  \bibfield  {author} {\bibinfo {author} {\bibfnamefont {P.~R.}\ \bibnamefont {Saulson}},\ }\href {https://doi.org/10.1103/PhysRevD.42.2437} {\bibfield  {journal} {\bibinfo  {journal} {Phys. Rev. D}\ }\textbf {\bibinfo {volume} {42}},\ \bibinfo {pages} {2437} (\bibinfo {year} {1990})}\BibitemShut {NoStop}%
\bibitem [{\citenamefont {Dobrindt}\ \emph {et~al.}(2008)\citenamefont {Dobrindt}, \citenamefont {{Wilson-Rae}},\ and\ \citenamefont {Kippenberg}}]{dobrindt2008ParametricNormalModeSplitting}%
  \BibitemOpen
  \bibfield  {author} {\bibinfo {author} {\bibfnamefont {J.~M.}\ \bibnamefont {Dobrindt}}, \bibinfo {author} {\bibfnamefont {I.}~\bibnamefont {{Wilson-Rae}}},\ and\ \bibinfo {author} {\bibfnamefont {T.~J.}\ \bibnamefont {Kippenberg}},\ }\href {https://doi.org/10.1103/PhysRevLett.101.263602} {\bibfield  {journal} {\bibinfo  {journal} {Phys. Rev. Lett.}\ }\textbf {\bibinfo {volume} {101}},\ \bibinfo {pages} {263602} (\bibinfo {year} {2008})}\BibitemShut {NoStop}%
\bibitem [{\citenamefont {Thompson}\ \emph {et~al.}(1992)\citenamefont {Thompson}, \citenamefont {Rempe},\ and\ \citenamefont {Kimble}}]{thompson1992ObservationNormalmodeSplitting}%
  \BibitemOpen
  \bibfield  {author} {\bibinfo {author} {\bibfnamefont {R.~J.}\ \bibnamefont {Thompson}}, \bibinfo {author} {\bibfnamefont {G.}~\bibnamefont {Rempe}},\ and\ \bibinfo {author} {\bibfnamefont {H.~J.}\ \bibnamefont {Kimble}},\ }\href {https://doi.org/10.1103/PhysRevLett.68.1132} {\bibfield  {journal} {\bibinfo  {journal} {Phys. Rev. Lett.}\ }\textbf {\bibinfo {volume} {68}},\ \bibinfo {pages} {1132} (\bibinfo {year} {1992})}\BibitemShut {NoStop}%
\bibitem [{\citenamefont {Frisk~Kockum}\ \emph {et~al.}(2019)\citenamefont {Frisk~Kockum}, \citenamefont {Miranowicz}, \citenamefont {De~Liberato}, \citenamefont {Savasta},\ and\ \citenamefont {Nori}}]{friskkockum2019UltrastrongCouplingLight}%
  \BibitemOpen
  \bibfield  {author} {\bibinfo {author} {\bibfnamefont {A.}~\bibnamefont {Frisk~Kockum}}, \bibinfo {author} {\bibfnamefont {A.}~\bibnamefont {Miranowicz}}, \bibinfo {author} {\bibfnamefont {S.}~\bibnamefont {De~Liberato}}, \bibinfo {author} {\bibfnamefont {S.}~\bibnamefont {Savasta}},\ and\ \bibinfo {author} {\bibfnamefont {F.}~\bibnamefont {Nori}},\ }\href {https://doi.org/10.1038/s42254-018-0006-2} {\bibfield  {journal} {\bibinfo  {journal} {Nat Rev Phys}\ }\textbf {\bibinfo {volume} {1}},\ \bibinfo {pages} {19} (\bibinfo {year} {2019})}\BibitemShut {NoStop}%
\bibitem [{\citenamefont {Aspelmeyer}\ \emph {et~al.}(2014)\citenamefont {Aspelmeyer}, \citenamefont {Kippenberg},\ and\ \citenamefont {Marquardt}}]{aspelmeyer2014CavityOptomechanics}%
  \BibitemOpen
  \bibfield  {author} {\bibinfo {author} {\bibfnamefont {M.}~\bibnamefont {Aspelmeyer}}, \bibinfo {author} {\bibfnamefont {T.~J.}\ \bibnamefont {Kippenberg}},\ and\ \bibinfo {author} {\bibfnamefont {F.}~\bibnamefont {Marquardt}},\ }\href {https://doi.org/10.1103/RevModPhys.86.1391} {\bibfield  {journal} {\bibinfo  {journal} {Rev. Mod. Phys.}\ }\textbf {\bibinfo {volume} {86}},\ \bibinfo {pages} {1391} (\bibinfo {year} {2014})}\BibitemShut {NoStop}%
\bibitem [{\citenamefont {Dobrindt}\ and\ \citenamefont {Kippenberg}(2010)}]{dobrindt2010TheoreticalAnalysisMechanical}%
  \BibitemOpen
  \bibfield  {author} {\bibinfo {author} {\bibfnamefont {J.~M.}\ \bibnamefont {Dobrindt}}\ and\ \bibinfo {author} {\bibfnamefont {T.~J.}\ \bibnamefont {Kippenberg}},\ }\href {https://doi.org/10.1103/PhysRevLett.104.033901} {\bibfield  {journal} {\bibinfo  {journal} {Phys. Rev. Lett.}\ }\textbf {\bibinfo {volume} {104}},\ \bibinfo {pages} {033901} (\bibinfo {year} {2010})}\BibitemShut {NoStop}%
\bibitem [{\citenamefont {Chan}\ \emph {et~al.}(2011)\citenamefont {Chan}, \citenamefont {Alegre}, \citenamefont {{Safavi-Naeini}}, \citenamefont {Hill}, \citenamefont {Krause}, \citenamefont {Gr{\"o}blacher}, \citenamefont {Aspelmeyer},\ and\ \citenamefont {Painter}}]{chan2011LaserCoolingNanomechanical}%
  \BibitemOpen
  \bibfield  {author} {\bibinfo {author} {\bibfnamefont {J.}~\bibnamefont {Chan}}, \bibinfo {author} {\bibfnamefont {T.~P.~M.}\ \bibnamefont {Alegre}}, \bibinfo {author} {\bibfnamefont {A.~H.}\ \bibnamefont {{Safavi-Naeini}}}, \bibinfo {author} {\bibfnamefont {J.~T.}\ \bibnamefont {Hill}}, \bibinfo {author} {\bibfnamefont {A.}~\bibnamefont {Krause}}, \bibinfo {author} {\bibfnamefont {S.}~\bibnamefont {Gr{\"o}blacher}}, \bibinfo {author} {\bibfnamefont {M.}~\bibnamefont {Aspelmeyer}},\ and\ \bibinfo {author} {\bibfnamefont {O.}~\bibnamefont {Painter}},\ }\href {https://doi.org/10.1038/nature10461} {\bibfield  {journal} {\bibinfo  {journal} {Nature}\ }\textbf {\bibinfo {volume} {478}},\ \bibinfo {pages} {89} (\bibinfo {year} {2011})}\BibitemShut {NoStop}%
\bibitem [{\citenamefont {M{\o}ller}\ \emph {et~al.}(2017)\citenamefont {M{\o}ller}, \citenamefont {Thomas}, \citenamefont {Vasilakis}, \citenamefont {Zeuthen}, \citenamefont {Tsaturyan}, \citenamefont {Balabas}, \citenamefont {Jensen}, \citenamefont {Schliesser}, \citenamefont {Hammerer},\ and\ \citenamefont {Polzik}}]{moller2017QuantumBackactionevadingMeasurement}%
  \BibitemOpen
  \bibfield  {author} {\bibinfo {author} {\bibfnamefont {C.~B.}\ \bibnamefont {M{\o}ller}}, \bibinfo {author} {\bibfnamefont {R.~A.}\ \bibnamefont {Thomas}}, \bibinfo {author} {\bibfnamefont {G.}~\bibnamefont {Vasilakis}}, \bibinfo {author} {\bibfnamefont {E.}~\bibnamefont {Zeuthen}}, \bibinfo {author} {\bibfnamefont {Y.}~\bibnamefont {Tsaturyan}}, \bibinfo {author} {\bibfnamefont {M.}~\bibnamefont {Balabas}}, \bibinfo {author} {\bibfnamefont {K.}~\bibnamefont {Jensen}}, \bibinfo {author} {\bibfnamefont {A.}~\bibnamefont {Schliesser}}, \bibinfo {author} {\bibfnamefont {K.}~\bibnamefont {Hammerer}},\ and\ \bibinfo {author} {\bibfnamefont {E.~S.}\ \bibnamefont {Polzik}},\ }\href {https://doi.org/10.1038/nature22980} {\bibfield  {journal} {\bibinfo  {journal} {Nature}\ }\textbf {\bibinfo {volume} {547}},\ \bibinfo {pages} {191} (\bibinfo {year} {2017})}\BibitemShut {NoStop}%
\bibitem [{\citenamefont {Hammerer}\ \emph {et~al.}(2010)\citenamefont {Hammerer}, \citenamefont {S{\o}rensen},\ and\ \citenamefont {Polzik}}]{hammerer2010QuantumInterfaceLight}%
  \BibitemOpen
  \bibfield  {author} {\bibinfo {author} {\bibfnamefont {K.}~\bibnamefont {Hammerer}}, \bibinfo {author} {\bibfnamefont {A.~S.}\ \bibnamefont {S{\o}rensen}},\ and\ \bibinfo {author} {\bibfnamefont {E.~S.}\ \bibnamefont {Polzik}},\ }\href {https://doi.org/10.1103/RevModPhys.82.1041} {\bibfield  {journal} {\bibinfo  {journal} {Rev. Mod. Phys.}\ }\textbf {\bibinfo {volume} {82}},\ \bibinfo {pages} {1041} (\bibinfo {year} {2010})}\BibitemShut {NoStop}%
\bibitem [{\citenamefont {Lu}\ \emph {et~al.}(2012)\citenamefont {Lu}, \citenamefont {Liu}, \citenamefont {Wang},\ and\ \citenamefont {Mao}}]{lu2012TunableHighchannelcountBandpass}%
  \BibitemOpen
  \bibfield  {author} {\bibinfo {author} {\bibfnamefont {H.}~\bibnamefont {Lu}}, \bibinfo {author} {\bibfnamefont {X.}~\bibnamefont {Liu}}, \bibinfo {author} {\bibfnamefont {G.}~\bibnamefont {Wang}},\ and\ \bibinfo {author} {\bibfnamefont {D.}~\bibnamefont {Mao}},\ }\href {https://doi.org/10.1088/0957-4484/23/44/444003} {\bibfield  {journal} {\bibinfo  {journal} {Nanotechnology}\ }\textbf {\bibinfo {volume} {23}},\ \bibinfo {pages} {444003} (\bibinfo {year} {2012})}\BibitemShut {NoStop}%
\bibitem [{\citenamefont {Smith}\ \emph {et~al.}(2004)\citenamefont {Smith}, \citenamefont {Chang}, \citenamefont {Fuller}, \citenamefont {Rosenberger},\ and\ \citenamefont {Boyd}}]{smith2004CoupledresonatorinducedTransparency}%
  \BibitemOpen
  \bibfield  {author} {\bibinfo {author} {\bibfnamefont {D.~D.}\ \bibnamefont {Smith}}, \bibinfo {author} {\bibfnamefont {H.}~\bibnamefont {Chang}}, \bibinfo {author} {\bibfnamefont {K.~A.}\ \bibnamefont {Fuller}}, \bibinfo {author} {\bibfnamefont {A.~T.}\ \bibnamefont {Rosenberger}},\ and\ \bibinfo {author} {\bibfnamefont {R.~W.}\ \bibnamefont {Boyd}},\ }\href {https://doi.org/10.1103/PhysRevA.69.063804} {\bibfield  {journal} {\bibinfo  {journal} {Phys. Rev. A}\ }\textbf {\bibinfo {volume} {69}},\ \bibinfo {pages} {063804} (\bibinfo {year} {2004})}\BibitemShut {NoStop}%
\bibitem [{\citenamefont {Zheng}\ \emph {et~al.}(2012)\citenamefont {Zheng}, \citenamefont {Jiang}, \citenamefont {Hua}, \citenamefont {Chang}, \citenamefont {Li}, \citenamefont {Fan},\ and\ \citenamefont {Xiao}}]{zheng2012ControllableOpticalAnaloga}%
  \BibitemOpen
  \bibfield  {author} {\bibinfo {author} {\bibfnamefont {C.}~\bibnamefont {Zheng}}, \bibinfo {author} {\bibfnamefont {X.}~\bibnamefont {Jiang}}, \bibinfo {author} {\bibfnamefont {S.}~\bibnamefont {Hua}}, \bibinfo {author} {\bibfnamefont {L.}~\bibnamefont {Chang}}, \bibinfo {author} {\bibfnamefont {G.}~\bibnamefont {Li}}, \bibinfo {author} {\bibfnamefont {H.}~\bibnamefont {Fan}},\ and\ \bibinfo {author} {\bibfnamefont {M.}~\bibnamefont {Xiao}},\ }\href {https://doi.org/10.1364/OE.20.018319} {\bibfield  {journal} {\bibinfo  {journal} {Opt. Express, OE}\ }\textbf {\bibinfo {volume} {20}},\ \bibinfo {pages} {18319} (\bibinfo {year} {2012})}\BibitemShut {NoStop}%
\bibitem [{\citenamefont {Guo}\ \emph {et~al.}(2021)\citenamefont {Guo}, \citenamefont {Zhang}, \citenamefont {Wu}, \citenamefont {Ye},\ and\ \citenamefont {Lin}}]{guo2021TransitionCoupledresonatorinducedTransparencya}%
  \BibitemOpen
  \bibfield  {author} {\bibinfo {author} {\bibfnamefont {S.-T.}\ \bibnamefont {Guo}}, \bibinfo {author} {\bibfnamefont {Y.-H.}\ \bibnamefont {Zhang}}, \bibinfo {author} {\bibfnamefont {L.-L.}\ \bibnamefont {Wu}}, \bibinfo {author} {\bibfnamefont {M.-Y.}\ \bibnamefont {Ye}},\ and\ \bibinfo {author} {\bibfnamefont {X.-M.}\ \bibnamefont {Lin}},\ }\href {https://doi.org/10.1103/PhysRevA.103.033510} {\bibfield  {journal} {\bibinfo  {journal} {Phys. Rev. A}\ }\textbf {\bibinfo {volume} {103}},\ \bibinfo {pages} {033510} (\bibinfo {year} {2021})}\BibitemShut {NoStop}%
\bibitem [{\citenamefont {Righini}\ and\ \citenamefont {Soria}(2016)}]{righini2016BiosensingWGMMicrospherical}%
  \BibitemOpen
  \bibfield  {author} {\bibinfo {author} {\bibfnamefont {G.~C.}\ \bibnamefont {Righini}}\ and\ \bibinfo {author} {\bibfnamefont {S.}~\bibnamefont {Soria}},\ }\href {https://doi.org/10.3390/s16060905} {\bibfield  {journal} {\bibinfo  {journal} {Sensors}\ }\textbf {\bibinfo {volume} {16}},\ \bibinfo {pages} {905} (\bibinfo {year} {2016})}\BibitemShut {NoStop}%
\bibitem [{\citenamefont {Totsuka}\ \emph {et~al.}(2007)\citenamefont {Totsuka}, \citenamefont {Kobayashi},\ and\ \citenamefont {Tomita}}]{totsuka2007SlowLightCoupledResonatorInduced}%
  \BibitemOpen
  \bibfield  {author} {\bibinfo {author} {\bibfnamefont {K.}~\bibnamefont {Totsuka}}, \bibinfo {author} {\bibfnamefont {N.}~\bibnamefont {Kobayashi}},\ and\ \bibinfo {author} {\bibfnamefont {M.}~\bibnamefont {Tomita}},\ }\href {https://doi.org/10.1103/PhysRevLett.98.213904} {\bibfield  {journal} {\bibinfo  {journal} {Phys. Rev. Lett.}\ }\textbf {\bibinfo {volume} {98}},\ \bibinfo {pages} {213904} (\bibinfo {year} {2007})}\BibitemShut {NoStop}%
\bibitem [{\citenamefont {Di}\ \emph {et~al.}(2011)\citenamefont {Di}, \citenamefont {Xie},\ and\ \citenamefont {Zhang}}]{di2011CoupledResonatorInducedTransparencySqueezed}%
  \BibitemOpen
  \bibfield  {author} {\bibinfo {author} {\bibfnamefont {K.}~\bibnamefont {Di}}, \bibinfo {author} {\bibfnamefont {C.}~\bibnamefont {Xie}},\ and\ \bibinfo {author} {\bibfnamefont {J.}~\bibnamefont {Zhang}},\ }\href {https://doi.org/10.1103/PhysRevLett.106.153602} {\bibfield  {journal} {\bibinfo  {journal} {Phys. Rev. Lett.}\ }\textbf {\bibinfo {volume} {106}},\ \bibinfo {pages} {153602} (\bibinfo {year} {2011})}\BibitemShut {NoStop}%
\bibitem [{\citenamefont {Li}\ \emph {et~al.}(2010)\citenamefont {Li}, \citenamefont {Wang}, \citenamefont {Su}, \citenamefont {Yan},\ and\ \citenamefont {Qiu}}]{li2010CoupledModeTheory}%
  \BibitemOpen
  \bibfield  {author} {\bibinfo {author} {\bibfnamefont {Q.}~\bibnamefont {Li}}, \bibinfo {author} {\bibfnamefont {T.}~\bibnamefont {Wang}}, \bibinfo {author} {\bibfnamefont {Y.}~\bibnamefont {Su}}, \bibinfo {author} {\bibfnamefont {M.}~\bibnamefont {Yan}},\ and\ \bibinfo {author} {\bibfnamefont {M.}~\bibnamefont {Qiu}},\ }\href {https://doi.org/10.1364/OE.18.008367} {\bibfield  {journal} {\bibinfo  {journal} {Opt. Express, OE}\ }\textbf {\bibinfo {volume} {18}},\ \bibinfo {pages} {8367} (\bibinfo {year} {2010})}\BibitemShut {NoStop}%
\bibitem [{\citenamefont {Korobko}\ \emph {et~al.}(2023{\natexlab{a}})\citenamefont {Korobko}, \citenamefont {S{\"u}dbeck}, \citenamefont {Steinlechner},\ and\ \citenamefont {Schnabel}}]{korobko2023FundamentalSensitivityLimit}%
  \BibitemOpen
  \bibfield  {author} {\bibinfo {author} {\bibfnamefont {M.}~\bibnamefont {Korobko}}, \bibinfo {author} {\bibfnamefont {J.}~\bibnamefont {S{\"u}dbeck}}, \bibinfo {author} {\bibfnamefont {S.}~\bibnamefont {Steinlechner}},\ and\ \bibinfo {author} {\bibfnamefont {R.}~\bibnamefont {Schnabel}},\ }\href {https://doi.org/10.1103/PhysRevA.108.063705} {\bibfield  {journal} {\bibinfo  {journal} {Phys. Rev. A}\ }\textbf {\bibinfo {volume} {108}},\ \bibinfo {pages} {063705} (\bibinfo {year} {2023}{\natexlab{a}})}\BibitemShut {NoStop}%
\bibitem [{\citenamefont {Gardner}\ \emph {et~al.}(2022)\citenamefont {Gardner}, \citenamefont {Yap}, \citenamefont {Adya}, \citenamefont {Chua}, \citenamefont {Slagmolen},\ and\ \citenamefont {McClelland}}]{gardner2022NondegenerateInternalSqueezinga}%
  \BibitemOpen
  \bibfield  {author} {\bibinfo {author} {\bibfnamefont {J.~W.}\ \bibnamefont {Gardner}}, \bibinfo {author} {\bibfnamefont {M.~J.}\ \bibnamefont {Yap}}, \bibinfo {author} {\bibfnamefont {V.}~\bibnamefont {Adya}}, \bibinfo {author} {\bibfnamefont {S.}~\bibnamefont {Chua}}, \bibinfo {author} {\bibfnamefont {B.~J.~J.}\ \bibnamefont {Slagmolen}},\ and\ \bibinfo {author} {\bibfnamefont {D.~E.}\ \bibnamefont {McClelland}},\ }\href {https://doi.org/10.1103/PhysRevD.106.L041101} {\bibfield  {journal} {\bibinfo  {journal} {Phys. Rev. D}\ }\textbf {\bibinfo {volume} {106}},\ \bibinfo {pages} {L041101} (\bibinfo {year} {2022})}\BibitemShut {NoStop}%
\bibitem [{\citenamefont {Somiya}\ \emph {et~al.}(2016)\citenamefont {Somiya}, \citenamefont {Kataoka}, \citenamefont {Kato}, \citenamefont {Saito},\ and\ \citenamefont {Yano}}]{somiya2016ParametricSignalAmplificationa}%
  \BibitemOpen
  \bibfield  {author} {\bibinfo {author} {\bibfnamefont {K.}~\bibnamefont {Somiya}}, \bibinfo {author} {\bibfnamefont {Y.}~\bibnamefont {Kataoka}}, \bibinfo {author} {\bibfnamefont {J.}~\bibnamefont {Kato}}, \bibinfo {author} {\bibfnamefont {N.}~\bibnamefont {Saito}},\ and\ \bibinfo {author} {\bibfnamefont {K.}~\bibnamefont {Yano}},\ }\href {https://doi.org/10.1016/j.physleta.2015.11.010} {\bibfield  {journal} {\bibinfo  {journal} {Physics Letters A}\ }\textbf {\bibinfo {volume} {380}},\ \bibinfo {pages} {521} (\bibinfo {year} {2016})}\BibitemShut {NoStop}%
\bibitem [{\citenamefont {Miao}\ \emph {et~al.}(2019)\citenamefont {Miao}, \citenamefont {Smith},\ and\ \citenamefont {Evans}}]{miao2019QuantumLimitLaser}%
  \BibitemOpen
  \bibfield  {author} {\bibinfo {author} {\bibfnamefont {H.}~\bibnamefont {Miao}}, \bibinfo {author} {\bibfnamefont {N.~D.}\ \bibnamefont {Smith}},\ and\ \bibinfo {author} {\bibfnamefont {M.}~\bibnamefont {Evans}},\ }\href {https://doi.org/10.1103/PhysRevX.9.011053} {\bibfield  {journal} {\bibinfo  {journal} {Phys. Rev. X}\ }\textbf {\bibinfo {volume} {9}},\ \bibinfo {pages} {011053} (\bibinfo {year} {2019})}\BibitemShut {NoStop}%
\bibitem [{\citenamefont {Rehbein}\ \emph {et~al.}(2005)\citenamefont {Rehbein}, \citenamefont {Harms}, \citenamefont {Schnabel},\ and\ \citenamefont {Danzmann}}]{rehbein2005OpticalTransferFunctions}%
  \BibitemOpen
  \bibfield  {author} {\bibinfo {author} {\bibfnamefont {H.}~\bibnamefont {Rehbein}}, \bibinfo {author} {\bibfnamefont {J.}~\bibnamefont {Harms}}, \bibinfo {author} {\bibfnamefont {R.}~\bibnamefont {Schnabel}},\ and\ \bibinfo {author} {\bibfnamefont {K.}~\bibnamefont {Danzmann}},\ }\href {https://doi.org/10.1103/PhysRevLett.95.193001} {\bibfield  {journal} {\bibinfo  {journal} {Phys. Rev. Lett.}\ }\textbf {\bibinfo {volume} {95}},\ \bibinfo {pages} {193001} (\bibinfo {year} {2005})}\BibitemShut {NoStop}%
\bibitem [{\citenamefont {Peano}\ \emph {et~al.}(2015)\citenamefont {Peano}, \citenamefont {Schwefel}, \citenamefont {Marquardt},\ and\ \citenamefont {Marquardt}}]{peano2015IntracavitySqueezingCan}%
  \BibitemOpen
  \bibfield  {author} {\bibinfo {author} {\bibfnamefont {V.}~\bibnamefont {Peano}}, \bibinfo {author} {\bibfnamefont {H.~G.~L.}\ \bibnamefont {Schwefel}}, \bibinfo {author} {\bibfnamefont {{\relax Ch}.}~\bibnamefont {Marquardt}},\ and\ \bibinfo {author} {\bibfnamefont {F.}~\bibnamefont {Marquardt}},\ }\href {https://doi.org/10.1103/PhysRevLett.115.243603} {\bibfield  {journal} {\bibinfo  {journal} {Phys. Rev. Lett.}\ }\textbf {\bibinfo {volume} {115}},\ \bibinfo {pages} {243603} (\bibinfo {year} {2015})}\BibitemShut {NoStop}%
\bibitem [{\citenamefont {Korobko}\ \emph {et~al.}(2017)\citenamefont {Korobko}, \citenamefont {Kleybolte}, \citenamefont {Ast}, \citenamefont {Miao}, \citenamefont {Chen},\ and\ \citenamefont {Schnabel}}]{korobko2017BeatingStandardSensitivityBandwidth}%
  \BibitemOpen
  \bibfield  {author} {\bibinfo {author} {\bibfnamefont {M.}~\bibnamefont {Korobko}}, \bibinfo {author} {\bibfnamefont {L.}~\bibnamefont {Kleybolte}}, \bibinfo {author} {\bibfnamefont {S.}~\bibnamefont {Ast}}, \bibinfo {author} {\bibfnamefont {H.}~\bibnamefont {Miao}}, \bibinfo {author} {\bibfnamefont {Y.}~\bibnamefont {Chen}},\ and\ \bibinfo {author} {\bibfnamefont {R.}~\bibnamefont {Schnabel}},\ }\href {https://doi.org/10.1103/PhysRevLett.118.143601} {\bibfield  {journal} {\bibinfo  {journal} {Phys. Rev. Lett.}\ }\textbf {\bibinfo {volume} {118}},\ \bibinfo {pages} {143601} (\bibinfo {year} {2017})}\BibitemShut {NoStop}%
\bibitem [{\citenamefont {Korobko}\ \emph {et~al.}(2023{\natexlab{b}})\citenamefont {Korobko}, \citenamefont {S{\"u}dbeck}, \citenamefont {Steinlechner},\ and\ \citenamefont {Schnabel}}]{korobko2023MitigatingQuantumDecoherence}%
  \BibitemOpen
  \bibfield  {author} {\bibinfo {author} {\bibfnamefont {M.}~\bibnamefont {Korobko}}, \bibinfo {author} {\bibfnamefont {J.}~\bibnamefont {S{\"u}dbeck}}, \bibinfo {author} {\bibfnamefont {S.}~\bibnamefont {Steinlechner}},\ and\ \bibinfo {author} {\bibfnamefont {R.}~\bibnamefont {Schnabel}},\ }\href {https://doi.org/10.1103/PhysRevLett.131.143603} {\bibfield  {journal} {\bibinfo  {journal} {Phys. Rev. Lett.}\ }\textbf {\bibinfo {volume} {131}},\ \bibinfo {pages} {143603} (\bibinfo {year} {2023}{\natexlab{b}})}\BibitemShut {NoStop}%
\bibitem [{\citenamefont {B{\"o}ttner}\ \emph {et~al.}(2024)\citenamefont {B{\"o}ttner}, \citenamefont {Bentley}, \citenamefont {Schnabel},\ and\ \citenamefont {Korobko}}]{bottner2024CoherentFeedbackQuantum}%
  \BibitemOpen
  \bibfield  {author} {\bibinfo {author} {\bibfnamefont {N.}~\bibnamefont {B{\"o}ttner}}, \bibinfo {author} {\bibfnamefont {J.}~\bibnamefont {Bentley}}, \bibinfo {author} {\bibfnamefont {R.}~\bibnamefont {Schnabel}},\ and\ \bibinfo {author} {\bibfnamefont {M.}~\bibnamefont {Korobko}},\ }\href {https://doi.org/10.1103/PhysRevD.110.103010} {\bibfield  {journal} {\bibinfo  {journal} {Phys. Rev. D}\ }\textbf {\bibinfo {volume} {110}},\ \bibinfo {pages} {103010} (\bibinfo {year} {2024})}\BibitemShut {NoStop}%
\bibitem [{\citenamefont {Korobko}\ \emph {et~al.}(2019)\citenamefont {Korobko}, \citenamefont {Ma}, \citenamefont {Chen},\ and\ \citenamefont {Schnabel}}]{korobko2019QuantumExpanderGravitationalwave}%
  \BibitemOpen
  \bibfield  {author} {\bibinfo {author} {\bibfnamefont {M.}~\bibnamefont {Korobko}}, \bibinfo {author} {\bibfnamefont {Y.}~\bibnamefont {Ma}}, \bibinfo {author} {\bibfnamefont {Y.}~\bibnamefont {Chen}},\ and\ \bibinfo {author} {\bibfnamefont {R.}~\bibnamefont {Schnabel}},\ }\href {https://doi.org/10.1038/s41377-019-0230-2} {\bibfield  {journal} {\bibinfo  {journal} {Light Sci Appl}\ }\textbf {\bibinfo {volume} {8}},\ \bibinfo {pages} {118} (\bibinfo {year} {2019})}\BibitemShut {NoStop}%
\bibitem [{\citenamefont {Adya}\ \emph {et~al.}(2020)\citenamefont {Adya}, \citenamefont {Yap}, \citenamefont {T{\"o}yr{\"a}}, \citenamefont {McRae}, \citenamefont {Altin}, \citenamefont {Sarre}, \citenamefont {Meijerink}, \citenamefont {Kijbunchoo}, \citenamefont {Slagmolen}, \citenamefont {Ward},\ and\ \citenamefont {McClelland}}]{adya2020QuantumEnhancedKHz}%
  \BibitemOpen
  \bibfield  {author} {\bibinfo {author} {\bibfnamefont {V.~B.}\ \bibnamefont {Adya}}, \bibinfo {author} {\bibfnamefont {M.~J.}\ \bibnamefont {Yap}}, \bibinfo {author} {\bibfnamefont {D.}~\bibnamefont {T{\"o}yr{\"a}}}, \bibinfo {author} {\bibfnamefont {T.~G.}\ \bibnamefont {McRae}}, \bibinfo {author} {\bibfnamefont {P.~A.}\ \bibnamefont {Altin}}, \bibinfo {author} {\bibfnamefont {L.~K.}\ \bibnamefont {Sarre}}, \bibinfo {author} {\bibfnamefont {M.}~\bibnamefont {Meijerink}}, \bibinfo {author} {\bibfnamefont {N.}~\bibnamefont {Kijbunchoo}}, \bibinfo {author} {\bibfnamefont {B.~J.~J.}\ \bibnamefont {Slagmolen}}, \bibinfo {author} {\bibfnamefont {R.~L.}\ \bibnamefont {Ward}},\ and\ \bibinfo {author} {\bibfnamefont {D.~E.}\ \bibnamefont {McClelland}},\ }\href {https://doi.org/10.1088/1361-6382/ab7615} {\bibfield  {journal} {\bibinfo  {journal} {Class. Quantum Grav.}\ }\textbf {\bibinfo {volume} {37}},\ \bibinfo {pages} {07LT02} (\bibinfo {year} {2020})}\BibitemShut {NoStop}%
\bibitem [{\citenamefont {Lv}\ \emph {et~al.}(2022)\citenamefont {Lv}, \citenamefont {Wang},\ and\ \citenamefont {Wang}}]{lv2022GainEnhancedSecond}%
  \BibitemOpen
  \bibfield  {author} {\bibinfo {author} {\bibfnamefont {X.-X.}\ \bibnamefont {Lv}}, \bibinfo {author} {\bibfnamefont {T.-J.}\ \bibnamefont {Wang}},\ and\ \bibinfo {author} {\bibfnamefont {C.}~\bibnamefont {Wang}},\ }\href {https://doi.org/10.1007/s10773-022-04977-3} {\bibfield  {journal} {\bibinfo  {journal} {Int J Theor Phys}\ }\textbf {\bibinfo {volume} {61}},\ \bibinfo {pages} {3} (\bibinfo {year} {2022})}\BibitemShut {NoStop}%
\bibitem [{\citenamefont {Huang}\ and\ \citenamefont {Batelaan}(2017)}]{huang2017DualismOpticalDifference}%
  \BibitemOpen
  \bibfield  {author} {\bibinfo {author} {\bibfnamefont {W.~C.-W.}\ \bibnamefont {Huang}}\ and\ \bibinfo {author} {\bibfnamefont {H.}~\bibnamefont {Batelaan}},\ }\href {https://doi.org/10.1209/0295-5075/119/24002} {\bibfield  {journal} {\bibinfo  {journal} {EPL}\ }\textbf {\bibinfo {volume} {119}},\ \bibinfo {pages} {24002} (\bibinfo {year} {2017})}\BibitemShut {NoStop}%
\bibitem [{\citenamefont {Liu}\ \emph {et~al.}(2020)\citenamefont {Liu}, \citenamefont {Yang}, \citenamefont {Liu},\ and\ \citenamefont {Zhu}}]{liu2020PhotonBlockadeEnhancing}%
  \BibitemOpen
  \bibfield  {author} {\bibinfo {author} {\bibfnamefont {J.-S.}\ \bibnamefont {Liu}}, \bibinfo {author} {\bibfnamefont {J.-Y.}\ \bibnamefont {Yang}}, \bibinfo {author} {\bibfnamefont {H.-Y.}\ \bibnamefont {Liu}},\ and\ \bibinfo {author} {\bibfnamefont {A.-D.}\ \bibnamefont {Zhu}},\ }\href {https://doi.org/10.1364/OE.395618} {\bibfield  {journal} {\bibinfo  {journal} {Opt. Express}\ }\textbf {\bibinfo {volume} {28}},\ \bibinfo {pages} {18397} (\bibinfo {year} {2020})}\BibitemShut {NoStop}%
\bibitem [{\citenamefont {Jabri}\ and\ \citenamefont {Eleuch}(2024)}]{jabri2024LightSqueezingEnhancement}%
  \BibitemOpen
  \bibfield  {author} {\bibinfo {author} {\bibfnamefont {H.}~\bibnamefont {Jabri}}\ and\ \bibinfo {author} {\bibfnamefont {H.}~\bibnamefont {Eleuch}},\ }\href {https://doi.org/10.1038/s41598-024-58447-3} {\bibfield  {journal} {\bibinfo  {journal} {Sci Rep}\ }\textbf {\bibinfo {volume} {14}},\ \bibinfo {pages} {7753} (\bibinfo {year} {2024})}\BibitemShut {NoStop}%
\bibitem [{\citenamefont {Walls}\ and\ \citenamefont {Milburn}(2008)}]{walls2008QuantumOptics}%
  \BibitemOpen
  \bibfield  {author} {\bibinfo {author} {\bibfnamefont {D.~F.}\ \bibnamefont {Walls}}\ and\ \bibinfo {author} {\bibfnamefont {G.~J.}\ \bibnamefont {Milburn}},\ }\href {https://doi.org/10.1007/978-3-540-28574-8} {\emph {\bibinfo {title} {Quantum {{Optics}}}}},\ \bibinfo {edition} {2nd}\ ed.,\ Springer {{eBook Collection}}\ (\bibinfo  {publisher} {Springer Berlin Heidelberg},\ \bibinfo {address} {Berlin, Heidelberg},\ \bibinfo {year} {2008})\BibitemShut {NoStop}%
\bibitem [{\citenamefont {Collett}\ and\ \citenamefont {Gardiner}(1984)}]{collett1984SqueezingIntracavityTravelingwavea}%
  \BibitemOpen
  \bibfield  {author} {\bibinfo {author} {\bibfnamefont {M.~J.}\ \bibnamefont {Collett}}\ and\ \bibinfo {author} {\bibfnamefont {C.~W.}\ \bibnamefont {Gardiner}},\ }\href {https://doi.org/10.1103/PhysRevA.30.1386} {\bibfield  {journal} {\bibinfo  {journal} {Phys. Rev. A}\ }\textbf {\bibinfo {volume} {30}},\ \bibinfo {pages} {1386} (\bibinfo {year} {1984})}\BibitemShut {NoStop}%
\bibitem [{\citenamefont {Collett}\ and\ \citenamefont {Walls}(1985)}]{collett1985SqueezingSpectraNonlinear}%
  \BibitemOpen
  \bibfield  {author} {\bibinfo {author} {\bibfnamefont {M.~J.}\ \bibnamefont {Collett}}\ and\ \bibinfo {author} {\bibfnamefont {D.~F.}\ \bibnamefont {Walls}},\ }\href {https://doi.org/10.1103/PhysRevA.32.2887} {\bibfield  {journal} {\bibinfo  {journal} {Phys. Rev. A}\ }\textbf {\bibinfo {volume} {32}},\ \bibinfo {pages} {2887} (\bibinfo {year} {1985})}\BibitemShut {NoStop}%
\bibitem [{\citenamefont {{de Vine}}\ \emph {et~al.}(2003)\citenamefont {{de Vine}}, \citenamefont {Gray}, \citenamefont {McClelland}, \citenamefont {Chen},\ and\ \citenamefont {Whitcomb}}]{devine2003MeasurementFrequencyResponse}%
  \BibitemOpen
  \bibfield  {author} {\bibinfo {author} {\bibfnamefont {G.}~\bibnamefont {{de Vine}}}, \bibinfo {author} {\bibfnamefont {M.}~\bibnamefont {Gray}}, \bibinfo {author} {\bibfnamefont {D.~E.}\ \bibnamefont {McClelland}}, \bibinfo {author} {\bibfnamefont {Y.}~\bibnamefont {Chen}},\ and\ \bibinfo {author} {\bibfnamefont {S.}~\bibnamefont {Whitcomb}},\ }\href {https://doi.org/10.1016/S0375-9601(03)01102-2} {\bibfield  {journal} {\bibinfo  {journal} {Physics Letters A}\ }\textbf {\bibinfo {volume} {316}},\ \bibinfo {pages} {17} (\bibinfo {year} {2003})}\BibitemShut {NoStop}%
\bibitem [{\citenamefont {Maggiore}\ \emph {et~al.}(2024)\citenamefont {Maggiore}, \citenamefont {Freise}, \citenamefont {Dmitriev},\ and\ \citenamefont {Sall{\'e}}}]{maggiore2024TuningResonantDoublets}%
  \BibitemOpen
  \bibfield  {author} {\bibinfo {author} {\bibfnamefont {R.}~\bibnamefont {Maggiore}}, \bibinfo {author} {\bibfnamefont {A.}~\bibnamefont {Freise}}, \bibinfo {author} {\bibfnamefont {A.}~\bibnamefont {Dmitriev}},\ and\ \bibinfo {author} {\bibfnamefont {M.}~\bibnamefont {Sall{\'e}}},\ }\href {https://doi.org/10.1103/PhysRevD.109.022010} {\bibfield  {journal} {\bibinfo  {journal} {Phys. Rev. D}\ }\textbf {\bibinfo {volume} {109}},\ \bibinfo {pages} {022010} (\bibinfo {year} {2024})}\BibitemShut {NoStop}%
\bibitem [{\citenamefont {Hansch}\ and\ \citenamefont {Couillaud}(1980)}]{hansch1980LaserFrequencyStabilization}%
  \BibitemOpen
  \bibfield  {author} {\bibinfo {author} {\bibfnamefont {T.~W.}\ \bibnamefont {Hansch}}\ and\ \bibinfo {author} {\bibfnamefont {B.}~\bibnamefont {Couillaud}},\ }\href {https://doi.org/10.1016/0030-4018(80)90069-3} {\bibfield  {journal} {\bibinfo  {journal} {Optics Communications}\ }\textbf {\bibinfo {volume} {35}},\ \bibinfo {pages} {441} (\bibinfo {year} {1980})}\BibitemShut {NoStop}%
\bibitem [{\citenamefont {Vahlbruch}\ \emph {et~al.}(2016)\citenamefont {Vahlbruch}, \citenamefont {Mehmet}, \citenamefont {Danzmann},\ and\ \citenamefont {Schnabel}}]{vahlbruch2016Detection15DB}%
  \BibitemOpen
  \bibfield  {author} {\bibinfo {author} {\bibfnamefont {H.}~\bibnamefont {Vahlbruch}}, \bibinfo {author} {\bibfnamefont {M.}~\bibnamefont {Mehmet}}, \bibinfo {author} {\bibfnamefont {K.}~\bibnamefont {Danzmann}},\ and\ \bibinfo {author} {\bibfnamefont {R.}~\bibnamefont {Schnabel}},\ }\href {https://doi.org/10.1103/PhysRevLett.117.110801} {\bibfield  {journal} {\bibinfo  {journal} {Phys. Rev. Lett.}\ }\textbf {\bibinfo {volume} {117}},\ \bibinfo {pages} {110801} (\bibinfo {year} {2016})}\BibitemShut {NoStop}%
\bibitem [{\citenamefont {{The LIGO Scientific Collaboration}}(2015)}]{collaboration2015AdvancedLIGO}%
  \BibitemOpen
  \bibfield  {author} {\bibinfo {author} {\bibnamefont {{The LIGO Scientific Collaboration}}},\ }\href {https://doi.org/10.1088/0264-9381/32/7/074001} {\bibfield  {journal} {\bibinfo  {journal} {Class. Quantum Grav.}\ }\textbf {\bibinfo {volume} {32}},\ \bibinfo {pages} {074001} (\bibinfo {year} {2015})}\BibitemShut {NoStop}%
\bibitem [{\citenamefont {{LIGO O4 Detector Collaboration}}\ \emph {et~al.}(2023)\citenamefont {{LIGO O4 Detector Collaboration}}, \citenamefont {Ganapathy}, \citenamefont {Jia}, \citenamefont {Nakano}, \citenamefont {Xu}, \citenamefont {Aritomi}, \citenamefont {Cullen}, \citenamefont {Kijbunchoo}, \citenamefont {Dwyer}, \citenamefont {Mullavey}, \citenamefont {McCuller}, \citenamefont {Abbott}, \citenamefont {Abouelfettouh}, \citenamefont {Adhikari}, \citenamefont {Ananyeva}, \citenamefont {Appert}, \citenamefont {Arai}, \citenamefont {Aston}, \citenamefont {Ball}, \citenamefont {Ballmer}, \citenamefont {Barker}, \citenamefont {Barsotti}, \citenamefont {Berger}, \citenamefont {Betzwieser}, \citenamefont {Bhattacharjee}, \citenamefont {Billingsley}, \citenamefont {Biscans}, \citenamefont {Bode}, \citenamefont {Bonilla}, \citenamefont {Bossilkov}, \citenamefont {Branch}, \citenamefont {Brooks}, \citenamefont {Brown}, \citenamefont {Bryant}, \citenamefont {Cahillane}, \citenamefont {Cao}, \citenamefont
  {Capote}, \citenamefont {Clara}, \citenamefont {Collins}, \citenamefont {Compton}, \citenamefont {Cottingham}, \citenamefont {Coyne}, \citenamefont {Crouch}, \citenamefont {Csizmazia}, \citenamefont {Dartez}, \citenamefont {Demos}, \citenamefont {Dohmen}, \citenamefont {Driggers}, \citenamefont {Effler}, \citenamefont {Ejlli}, \citenamefont {Etzel}, \citenamefont {Evans}, \citenamefont {Feicht}, \citenamefont {Frey}, \citenamefont {Frischhertz}, \citenamefont {Fritschel}, \citenamefont {Frolov}, \citenamefont {Fulda}, \citenamefont {Fyffe}, \citenamefont {Gateley}, \citenamefont {Giaime}, \citenamefont {Giardina}, \citenamefont {Glanzer}, \citenamefont {Goetz}, \citenamefont {Goetz}, \citenamefont {{Goodwin-Jones}}, \citenamefont {Gras}, \citenamefont {Gray}, \citenamefont {Griffith}, \citenamefont {Grote}, \citenamefont {Guidry}, \citenamefont {Hall}, \citenamefont {Hanks}, \citenamefont {Hanson}, \citenamefont {Heintze}, \citenamefont {{Helmling-Cornell}}, \citenamefont {Holland}, \citenamefont {Hoyland},
  \citenamefont {Huang}, \citenamefont {Inoue}, \citenamefont {James}, \citenamefont {Jennings}, \citenamefont {Karat}, \citenamefont {Karki}, \citenamefont {Kasprzack}, \citenamefont {Kawabe}, \citenamefont {King}, \citenamefont {Kissel}, \citenamefont {Komori}, \citenamefont {Kontos}, \citenamefont {Kumar}, \citenamefont {Kuns}, \citenamefont {Landry}, \citenamefont {Lantz}, \citenamefont {Laxen}, \citenamefont {Lee}, \citenamefont {Lesovsky}, \citenamefont {Llamas}, \citenamefont {Lormand}, \citenamefont {Loughlin}, \citenamefont {Macas}, \citenamefont {MacInnis}, \citenamefont {Makarem}, \citenamefont {Mannix}, \citenamefont {Mansell}, \citenamefont {Martin}, \citenamefont {Mason}, \citenamefont {Matichard}, \citenamefont {Mavalvala}, \citenamefont {Maxwell}, \citenamefont {McCarrol}, \citenamefont {McCarthy}, \citenamefont {McClelland}, \citenamefont {McCormick}, \citenamefont {McRae}, \citenamefont {Mera}, \citenamefont {Merilh}, \citenamefont {Meylahn}, \citenamefont {Mittleman}, \citenamefont
  {Moraru}, \citenamefont {Moreno}, \citenamefont {Nelson}, \citenamefont {Neunzert}, \citenamefont {Notte}, \citenamefont {Oberling}, \citenamefont {O'Hanlon}, \citenamefont {Osthelder}, \citenamefont {Ottaway}, \citenamefont {Overmier}, \citenamefont {Parker}, \citenamefont {Pele}, \citenamefont {Pham}, \citenamefont {Pirello}, \citenamefont {Quetschke}, \citenamefont {Ramirez}, \citenamefont {Reyes}, \citenamefont {Richardson}, \citenamefont {Robinson}, \citenamefont {Rollins}, \citenamefont {Romel}, \citenamefont {Romie}, \citenamefont {Ross}, \citenamefont {Ryan}, \citenamefont {Sadecki}, \citenamefont {Sanchez}, \citenamefont {Sanchez}, \citenamefont {Sanchez}, \citenamefont {Savage}, \citenamefont {Schaetzl}, \citenamefont {Schiworski}, \citenamefont {Schnabel}, \citenamefont {Schofield}, \citenamefont {Schwartz}, \citenamefont {Sellers}, \citenamefont {Shaffer}, \citenamefont {Short}, \citenamefont {Sigg}, \citenamefont {Slagmolen}, \citenamefont {Soike}, \citenamefont {Soni}, \citenamefont
  {Srivastava}, \citenamefont {Sun}, \citenamefont {Tanner}, \citenamefont {Thomas}, \citenamefont {Thomas}, \citenamefont {Thorne}, \citenamefont {Torrie}, \citenamefont {Traylor}, \citenamefont {Ubhi}, \citenamefont {Vajente}, \citenamefont {Vanosky}, \citenamefont {Vecchio}, \citenamefont {Veitch}, \citenamefont {Vibhute}, \citenamefont {{von Reis}}, \citenamefont {Warner}, \citenamefont {Weaver}, \citenamefont {Weiss}, \citenamefont {Whittle}, \citenamefont {Willke}, \citenamefont {Wipf}, \citenamefont {Yamamoto}, \citenamefont {Zhang},\ and\ \citenamefont {Zucker}}]{ligoo4detectorcollaboration2023BroadbandQuantumEnhancement}%
  \BibitemOpen
  \bibfield  {author} {\bibinfo {author} {\bibnamefont {{LIGO O4 Detector Collaboration}}}, \bibinfo {author} {\bibfnamefont {D.}~\bibnamefont {Ganapathy}}, \bibinfo {author} {\bibfnamefont {W.}~\bibnamefont {Jia}}, \bibinfo {author} {\bibfnamefont {M.}~\bibnamefont {Nakano}}, \bibinfo {author} {\bibfnamefont {V.}~\bibnamefont {Xu}}, \bibinfo {author} {\bibfnamefont {N.}~\bibnamefont {Aritomi}}, \bibinfo {author} {\bibfnamefont {T.}~\bibnamefont {Cullen}}, \bibinfo {author} {\bibfnamefont {N.}~\bibnamefont {Kijbunchoo}}, \bibinfo {author} {\bibfnamefont {S.~E.}\ \bibnamefont {Dwyer}}, \bibinfo {author} {\bibfnamefont {A.}~\bibnamefont {Mullavey}}, \bibinfo {author} {\bibfnamefont {L.}~\bibnamefont {McCuller}}, \bibinfo {author} {\bibfnamefont {R.}~\bibnamefont {Abbott}}, \bibinfo {author} {\bibfnamefont {I.}~\bibnamefont {Abouelfettouh}}, \bibinfo {author} {\bibfnamefont {R.~X.}\ \bibnamefont {Adhikari}}, \bibinfo {author} {\bibfnamefont {A.}~\bibnamefont {Ananyeva}}, \bibinfo {author} {\bibfnamefont
  {S.}~\bibnamefont {Appert}}, \bibinfo {author} {\bibfnamefont {K.}~\bibnamefont {Arai}}, \bibinfo {author} {\bibfnamefont {S.~M.}\ \bibnamefont {Aston}}, \bibinfo {author} {\bibfnamefont {M.}~\bibnamefont {Ball}}, \bibinfo {author} {\bibfnamefont {S.~W.}\ \bibnamefont {Ballmer}}, \bibinfo {author} {\bibfnamefont {D.}~\bibnamefont {Barker}}, \bibinfo {author} {\bibfnamefont {L.}~\bibnamefont {Barsotti}}, \bibinfo {author} {\bibfnamefont {B.~K.}\ \bibnamefont {Berger}}, \bibinfo {author} {\bibfnamefont {J.}~\bibnamefont {Betzwieser}}, \bibinfo {author} {\bibfnamefont {D.}~\bibnamefont {Bhattacharjee}}, \bibinfo {author} {\bibfnamefont {G.}~\bibnamefont {Billingsley}}, \bibinfo {author} {\bibfnamefont {S.}~\bibnamefont {Biscans}}, \bibinfo {author} {\bibfnamefont {N.}~\bibnamefont {Bode}}, \bibinfo {author} {\bibfnamefont {E.}~\bibnamefont {Bonilla}}, \bibinfo {author} {\bibfnamefont {V.}~\bibnamefont {Bossilkov}}, \bibinfo {author} {\bibfnamefont {A.}~\bibnamefont {Branch}}, \bibinfo {author} {\bibfnamefont
  {A.~F.}\ \bibnamefont {Brooks}}, \bibinfo {author} {\bibfnamefont {D.~D.}\ \bibnamefont {Brown}}, \bibinfo {author} {\bibfnamefont {J.}~\bibnamefont {Bryant}}, \bibinfo {author} {\bibfnamefont {C.}~\bibnamefont {Cahillane}}, \bibinfo {author} {\bibfnamefont {H.}~\bibnamefont {Cao}}, \bibinfo {author} {\bibfnamefont {E.}~\bibnamefont {Capote}}, \bibinfo {author} {\bibfnamefont {F.}~\bibnamefont {Clara}}, \bibinfo {author} {\bibfnamefont {J.}~\bibnamefont {Collins}}, \bibinfo {author} {\bibfnamefont {C.~M.}\ \bibnamefont {Compton}}, \bibinfo {author} {\bibfnamefont {R.}~\bibnamefont {Cottingham}}, \bibinfo {author} {\bibfnamefont {D.~C.}\ \bibnamefont {Coyne}}, \bibinfo {author} {\bibfnamefont {R.}~\bibnamefont {Crouch}}, \bibinfo {author} {\bibfnamefont {J.}~\bibnamefont {Csizmazia}}, \bibinfo {author} {\bibfnamefont {L.~P.}\ \bibnamefont {Dartez}}, \bibinfo {author} {\bibfnamefont {N.}~\bibnamefont {Demos}}, \bibinfo {author} {\bibfnamefont {E.}~\bibnamefont {Dohmen}}, \bibinfo {author} {\bibfnamefont
  {J.~C.}\ \bibnamefont {Driggers}}, \bibinfo {author} {\bibfnamefont {A.}~\bibnamefont {Effler}}, \bibinfo {author} {\bibfnamefont {A.}~\bibnamefont {Ejlli}}, \bibinfo {author} {\bibfnamefont {T.}~\bibnamefont {Etzel}}, \bibinfo {author} {\bibfnamefont {M.}~\bibnamefont {Evans}}, \bibinfo {author} {\bibfnamefont {J.}~\bibnamefont {Feicht}}, \bibinfo {author} {\bibfnamefont {R.}~\bibnamefont {Frey}}, \bibinfo {author} {\bibfnamefont {W.}~\bibnamefont {Frischhertz}}, \bibinfo {author} {\bibfnamefont {P.}~\bibnamefont {Fritschel}}, \bibinfo {author} {\bibfnamefont {V.~V.}\ \bibnamefont {Frolov}}, \bibinfo {author} {\bibfnamefont {P.}~\bibnamefont {Fulda}}, \bibinfo {author} {\bibfnamefont {M.}~\bibnamefont {Fyffe}}, \bibinfo {author} {\bibfnamefont {B.}~\bibnamefont {Gateley}}, \bibinfo {author} {\bibfnamefont {J.~A.}\ \bibnamefont {Giaime}}, \bibinfo {author} {\bibfnamefont {K.~D.}\ \bibnamefont {Giardina}}, \bibinfo {author} {\bibfnamefont {J.}~\bibnamefont {Glanzer}}, \bibinfo {author} {\bibfnamefont
  {E.}~\bibnamefont {Goetz}}, \bibinfo {author} {\bibfnamefont {R.}~\bibnamefont {Goetz}}, \bibinfo {author} {\bibfnamefont {A.~W.}\ \bibnamefont {{Goodwin-Jones}}}, \bibinfo {author} {\bibfnamefont {S.}~\bibnamefont {Gras}}, \bibinfo {author} {\bibfnamefont {C.}~\bibnamefont {Gray}}, \bibinfo {author} {\bibfnamefont {D.}~\bibnamefont {Griffith}}, \bibinfo {author} {\bibfnamefont {H.}~\bibnamefont {Grote}}, \bibinfo {author} {\bibfnamefont {T.}~\bibnamefont {Guidry}}, \bibinfo {author} {\bibfnamefont {E.~D.}\ \bibnamefont {Hall}}, \bibinfo {author} {\bibfnamefont {J.}~\bibnamefont {Hanks}}, \bibinfo {author} {\bibfnamefont {J.}~\bibnamefont {Hanson}}, \bibinfo {author} {\bibfnamefont {M.~C.}\ \bibnamefont {Heintze}}, \bibinfo {author} {\bibfnamefont {A.~F.}\ \bibnamefont {{Helmling-Cornell}}}, \bibinfo {author} {\bibfnamefont {N.~A.}\ \bibnamefont {Holland}}, \bibinfo {author} {\bibfnamefont {D.}~\bibnamefont {Hoyland}}, \bibinfo {author} {\bibfnamefont {H.~Y.}\ \bibnamefont {Huang}}, \bibinfo {author}
  {\bibfnamefont {Y.}~\bibnamefont {Inoue}}, \bibinfo {author} {\bibfnamefont {A.~L.}\ \bibnamefont {James}}, \bibinfo {author} {\bibfnamefont {A.}~\bibnamefont {Jennings}}, \bibinfo {author} {\bibfnamefont {S.}~\bibnamefont {Karat}}, \bibinfo {author} {\bibfnamefont {S.}~\bibnamefont {Karki}}, \bibinfo {author} {\bibfnamefont {M.}~\bibnamefont {Kasprzack}}, \bibinfo {author} {\bibfnamefont {K.}~\bibnamefont {Kawabe}}, \bibinfo {author} {\bibfnamefont {P.~J.}\ \bibnamefont {King}}, \bibinfo {author} {\bibfnamefont {J.~S.}\ \bibnamefont {Kissel}}, \bibinfo {author} {\bibfnamefont {K.}~\bibnamefont {Komori}}, \bibinfo {author} {\bibfnamefont {A.}~\bibnamefont {Kontos}}, \bibinfo {author} {\bibfnamefont {R.}~\bibnamefont {Kumar}}, \bibinfo {author} {\bibfnamefont {K.}~\bibnamefont {Kuns}}, \bibinfo {author} {\bibfnamefont {M.}~\bibnamefont {Landry}}, \bibinfo {author} {\bibfnamefont {B.}~\bibnamefont {Lantz}}, \bibinfo {author} {\bibfnamefont {M.}~\bibnamefont {Laxen}}, \bibinfo {author} {\bibfnamefont
  {K.}~\bibnamefont {Lee}}, \bibinfo {author} {\bibfnamefont {M.}~\bibnamefont {Lesovsky}}, \bibinfo {author} {\bibfnamefont {F.}~\bibnamefont {Llamas}}, \bibinfo {author} {\bibfnamefont {M.}~\bibnamefont {Lormand}}, \bibinfo {author} {\bibfnamefont {H.~A.}\ \bibnamefont {Loughlin}}, \bibinfo {author} {\bibfnamefont {R.}~\bibnamefont {Macas}}, \bibinfo {author} {\bibfnamefont {M.}~\bibnamefont {MacInnis}}, \bibinfo {author} {\bibfnamefont {C.~N.}\ \bibnamefont {Makarem}}, \bibinfo {author} {\bibfnamefont {B.}~\bibnamefont {Mannix}}, \bibinfo {author} {\bibfnamefont {G.~L.}\ \bibnamefont {Mansell}}, \bibinfo {author} {\bibfnamefont {R.~M.}\ \bibnamefont {Martin}}, \bibinfo {author} {\bibfnamefont {K.}~\bibnamefont {Mason}}, \bibinfo {author} {\bibfnamefont {F.}~\bibnamefont {Matichard}}, \bibinfo {author} {\bibfnamefont {N.}~\bibnamefont {Mavalvala}}, \bibinfo {author} {\bibfnamefont {N.}~\bibnamefont {Maxwell}}, \bibinfo {author} {\bibfnamefont {G.}~\bibnamefont {McCarrol}}, \bibinfo {author} {\bibfnamefont
  {R.}~\bibnamefont {McCarthy}}, \bibinfo {author} {\bibfnamefont {D.~E.}\ \bibnamefont {McClelland}}, \bibinfo {author} {\bibfnamefont {S.}~\bibnamefont {McCormick}}, \bibinfo {author} {\bibfnamefont {T.}~\bibnamefont {McRae}}, \bibinfo {author} {\bibfnamefont {F.}~\bibnamefont {Mera}}, \bibinfo {author} {\bibfnamefont {E.~L.}\ \bibnamefont {Merilh}}, \bibinfo {author} {\bibfnamefont {F.}~\bibnamefont {Meylahn}}, \bibinfo {author} {\bibfnamefont {R.}~\bibnamefont {Mittleman}}, \bibinfo {author} {\bibfnamefont {D.}~\bibnamefont {Moraru}}, \bibinfo {author} {\bibfnamefont {G.}~\bibnamefont {Moreno}}, \bibinfo {author} {\bibfnamefont {T.~J.~N.}\ \bibnamefont {Nelson}}, \bibinfo {author} {\bibfnamefont {A.}~\bibnamefont {Neunzert}}, \bibinfo {author} {\bibfnamefont {J.}~\bibnamefont {Notte}}, \bibinfo {author} {\bibfnamefont {J.}~\bibnamefont {Oberling}}, \bibinfo {author} {\bibfnamefont {T.}~\bibnamefont {O'Hanlon}}, \bibinfo {author} {\bibfnamefont {C.}~\bibnamefont {Osthelder}}, \bibinfo {author}
  {\bibfnamefont {D.~J.}\ \bibnamefont {Ottaway}}, \bibinfo {author} {\bibfnamefont {H.}~\bibnamefont {Overmier}}, \bibinfo {author} {\bibfnamefont {W.}~\bibnamefont {Parker}}, \bibinfo {author} {\bibfnamefont {A.}~\bibnamefont {Pele}}, \bibinfo {author} {\bibfnamefont {H.}~\bibnamefont {Pham}}, \bibinfo {author} {\bibfnamefont {M.}~\bibnamefont {Pirello}}, \bibinfo {author} {\bibfnamefont {V.}~\bibnamefont {Quetschke}}, \bibinfo {author} {\bibfnamefont {K.~E.}\ \bibnamefont {Ramirez}}, \bibinfo {author} {\bibfnamefont {J.}~\bibnamefont {Reyes}}, \bibinfo {author} {\bibfnamefont {J.~W.}\ \bibnamefont {Richardson}}, \bibinfo {author} {\bibfnamefont {M.}~\bibnamefont {Robinson}}, \bibinfo {author} {\bibfnamefont {J.~G.}\ \bibnamefont {Rollins}}, \bibinfo {author} {\bibfnamefont {C.~L.}\ \bibnamefont {Romel}}, \bibinfo {author} {\bibfnamefont {J.~H.}\ \bibnamefont {Romie}}, \bibinfo {author} {\bibfnamefont {M.~P.}\ \bibnamefont {Ross}}, \bibinfo {author} {\bibfnamefont {K.}~\bibnamefont {Ryan}}, \bibinfo
  {author} {\bibfnamefont {T.}~\bibnamefont {Sadecki}}, \bibinfo {author} {\bibfnamefont {A.}~\bibnamefont {Sanchez}}, \bibinfo {author} {\bibfnamefont {E.~J.}\ \bibnamefont {Sanchez}}, \bibinfo {author} {\bibfnamefont {L.~E.}\ \bibnamefont {Sanchez}}, \bibinfo {author} {\bibfnamefont {R.~L.}\ \bibnamefont {Savage}}, \bibinfo {author} {\bibfnamefont {D.}~\bibnamefont {Schaetzl}}, \bibinfo {author} {\bibfnamefont {M.~G.}\ \bibnamefont {Schiworski}}, \bibinfo {author} {\bibfnamefont {R.}~\bibnamefont {Schnabel}}, \bibinfo {author} {\bibfnamefont {R.~M.~S.}\ \bibnamefont {Schofield}}, \bibinfo {author} {\bibfnamefont {E.}~\bibnamefont {Schwartz}}, \bibinfo {author} {\bibfnamefont {D.}~\bibnamefont {Sellers}}, \bibinfo {author} {\bibfnamefont {T.}~\bibnamefont {Shaffer}}, \bibinfo {author} {\bibfnamefont {R.~W.}\ \bibnamefont {Short}}, \bibinfo {author} {\bibfnamefont {D.}~\bibnamefont {Sigg}}, \bibinfo {author} {\bibfnamefont {B.~J.~J.}\ \bibnamefont {Slagmolen}}, \bibinfo {author} {\bibfnamefont
  {C.}~\bibnamefont {Soike}}, \bibinfo {author} {\bibfnamefont {S.}~\bibnamefont {Soni}}, \bibinfo {author} {\bibfnamefont {V.}~\bibnamefont {Srivastava}}, \bibinfo {author} {\bibfnamefont {L.}~\bibnamefont {Sun}}, \bibinfo {author} {\bibfnamefont {D.~B.}\ \bibnamefont {Tanner}}, \bibinfo {author} {\bibfnamefont {M.}~\bibnamefont {Thomas}}, \bibinfo {author} {\bibfnamefont {P.}~\bibnamefont {Thomas}}, \bibinfo {author} {\bibfnamefont {K.~A.}\ \bibnamefont {Thorne}}, \bibinfo {author} {\bibfnamefont {C.~I.}\ \bibnamefont {Torrie}}, \bibinfo {author} {\bibfnamefont {G.}~\bibnamefont {Traylor}}, \bibinfo {author} {\bibfnamefont {A.~S.}\ \bibnamefont {Ubhi}}, \bibinfo {author} {\bibfnamefont {G.}~\bibnamefont {Vajente}}, \bibinfo {author} {\bibfnamefont {J.}~\bibnamefont {Vanosky}}, \bibinfo {author} {\bibfnamefont {A.}~\bibnamefont {Vecchio}}, \bibinfo {author} {\bibfnamefont {P.~J.}\ \bibnamefont {Veitch}}, \bibinfo {author} {\bibfnamefont {A.~M.}\ \bibnamefont {Vibhute}}, \bibinfo {author} {\bibfnamefont
  {E.~R.~G.}\ \bibnamefont {{von Reis}}}, \bibinfo {author} {\bibfnamefont {J.}~\bibnamefont {Warner}}, \bibinfo {author} {\bibfnamefont {B.}~\bibnamefont {Weaver}}, \bibinfo {author} {\bibfnamefont {R.}~\bibnamefont {Weiss}}, \bibinfo {author} {\bibfnamefont {C.}~\bibnamefont {Whittle}}, \bibinfo {author} {\bibfnamefont {B.}~\bibnamefont {Willke}}, \bibinfo {author} {\bibfnamefont {C.~C.}\ \bibnamefont {Wipf}}, \bibinfo {author} {\bibfnamefont {H.}~\bibnamefont {Yamamoto}}, \bibinfo {author} {\bibfnamefont {L.}~\bibnamefont {Zhang}},\ and\ \bibinfo {author} {\bibfnamefont {M.~E.}\ \bibnamefont {Zucker}},\ }\href {https://doi.org/10.1103/PhysRevX.13.041021} {\bibfield  {journal} {\bibinfo  {journal} {Phys. Rev. X}\ }\textbf {\bibinfo {volume} {13}},\ \bibinfo {pages} {041021} (\bibinfo {year} {2023})}\BibitemShut {NoStop}%
\bibitem [{\citenamefont {Acernese}\ \emph {et~al.}(2014)\citenamefont {Acernese}, \citenamefont {Agathos}, \citenamefont {Agatsuma}, \citenamefont {Aisa}, \citenamefont {Allemandou}, \citenamefont {Allocca}, \citenamefont {Amarni}, \citenamefont {Astone}, \citenamefont {Balestri}, \citenamefont {Ballardin}, \citenamefont {Barone}, \citenamefont {Baronick}, \citenamefont {Barsuglia}, \citenamefont {Basti}, \citenamefont {Basti}, \citenamefont {Bauer}, \citenamefont {Bavigadda}, \citenamefont {Bejger}, \citenamefont {Beker}, \citenamefont {Belczynski}, \citenamefont {Bersanetti}, \citenamefont {Bertolini}, \citenamefont {Bitossi}, \citenamefont {Bizouard}, \citenamefont {Bloemen}, \citenamefont {Blom}, \citenamefont {Boer}, \citenamefont {Bogaert}, \citenamefont {Bondi}, \citenamefont {Bondu}, \citenamefont {Bonelli}, \citenamefont {Bonnand}, \citenamefont {Boschi}, \citenamefont {Bosi}, \citenamefont {Bouedo}, \citenamefont {Bradaschia}, \citenamefont {Branchesi}, \citenamefont {Briant}, \citenamefont
  {Brillet}, \citenamefont {Brisson}, \citenamefont {Bulik}, \citenamefont {Bulten}, \citenamefont {Buskulic}, \citenamefont {Buy}, \citenamefont {Cagnoli}, \citenamefont {Calloni}, \citenamefont {Campeggi}, \citenamefont {Canuel}, \citenamefont {Carbognani}, \citenamefont {Cavalier}, \citenamefont {Cavalieri}, \citenamefont {Cella}, \citenamefont {Cesarini}, \citenamefont {Mottin}, \citenamefont {Chincarini}, \citenamefont {Chiummo}, \citenamefont {Chua}, \citenamefont {Cleva}, \citenamefont {Coccia}, \citenamefont {Cohadon}, \citenamefont {Colla}, \citenamefont {Colombini}, \citenamefont {Conte}, \citenamefont {Coulon}, \citenamefont {Cuoco}, \citenamefont {Dalmaz}, \citenamefont {D'Antonio}, \citenamefont {Dattilo}, \citenamefont {Davier}, \citenamefont {Day}, \citenamefont {Debreczeni}, \citenamefont {Degallaix}, \citenamefont {Del{\'e}glise}, \citenamefont {Pozzo}, \citenamefont {Dereli}, \citenamefont {Rosa}, \citenamefont {Fiore}, \citenamefont {Lieto}, \citenamefont {Virgilio}, \citenamefont {Doets},
  \citenamefont {Dolique}, \citenamefont {Drago}, \citenamefont {Ducrot}, \citenamefont {Endr{\H o}czi}, \citenamefont {Fafone}, \citenamefont {Farinon}, \citenamefont {Ferrante}, \citenamefont {Ferrini}, \citenamefont {Fidecaro}, \citenamefont {Fiori}, \citenamefont {Flaminio}, \citenamefont {Fournier}, \citenamefont {Franco}, \citenamefont {Frasca}, \citenamefont {Frasconi}, \citenamefont {Gammaitoni}, \citenamefont {Garufi}, \citenamefont {Gaspard}, \citenamefont {Gatto}, \citenamefont {Gemme}, \citenamefont {Gendre}, \citenamefont {Genin}, \citenamefont {Gennai}, \citenamefont {Ghosh}, \citenamefont {Giacobone}, \citenamefont {Giazotto}, \citenamefont {Gouaty}, \citenamefont {Granata}, \citenamefont {Greco}, \citenamefont {Groot}, \citenamefont {Guidi}, \citenamefont {Harms}, \citenamefont {Heidmann}, \citenamefont {Heitmann}, \citenamefont {Hello}, \citenamefont {Hemming}, \citenamefont {Hennes}, \citenamefont {Hofman}, \citenamefont {Jaranowski}, \citenamefont {Jonker}, \citenamefont {Kasprzack},
  \citenamefont {K{\'e}f{\'e}lian}, \citenamefont {Kowalska}, \citenamefont {Kraan}, \citenamefont {Kr{\'o}lak}, \citenamefont {Kutynia}, \citenamefont {Lazzaro}, \citenamefont {Leonardi}, \citenamefont {Leroy}, \citenamefont {Letendre}, \citenamefont {Li}, \citenamefont {Lieunard}, \citenamefont {Lorenzini}, \citenamefont {Loriette}, \citenamefont {Losurdo}, \citenamefont {Magazz{\`u}}, \citenamefont {Majorana}, \citenamefont {Maksimovic}, \citenamefont {Malvezzi}, \citenamefont {Man}, \citenamefont {Mangano}, \citenamefont {Mantovani}, \citenamefont {Marchesoni}, \citenamefont {Marion}, \citenamefont {Marque}, \citenamefont {Martelli}, \citenamefont {Martellini}, \citenamefont {Masserot}, \citenamefont {Meacher}, \citenamefont {Meidam}, \citenamefont {Mezzani}, \citenamefont {Michel}, \citenamefont {Milano}, \citenamefont {Minenkov}, \citenamefont {Moggi}, \citenamefont {Mohan}, \citenamefont {Montani}, \citenamefont {Morgado}, \citenamefont {Mours}, \citenamefont {Mul}, \citenamefont {Nagy}, \citenamefont
  {Nardecchia}, \citenamefont {Naticchioni}, \citenamefont {Nelemans}, \citenamefont {Neri}, \citenamefont {Neri}, \citenamefont {Nocera}, \citenamefont {Pacaud}, \citenamefont {Palomba}, \citenamefont {Paoletti}, \citenamefont {Paoli}, \citenamefont {Pasqualetti}, \citenamefont {Passaquieti}, \citenamefont {Passuello}, \citenamefont {Perciballi}, \citenamefont {Petit}, \citenamefont {Pichot}, \citenamefont {Piergiovanni}, \citenamefont {Pillant}, \citenamefont {Piluso}, \citenamefont {Pinard}, \citenamefont {Poggiani}, \citenamefont {Prijatelj}, \citenamefont {Prodi}, \citenamefont {Punturo}, \citenamefont {Puppo}, \citenamefont {Rabeling}, \citenamefont {R{\'a}cz}, \citenamefont {Rapagnani}, \citenamefont {Razzano}, \citenamefont {Re}, \citenamefont {Regimbau}, \citenamefont {Ricci}, \citenamefont {Robinet}, \citenamefont {Rocchi}, \citenamefont {Rolland}, \citenamefont {Romano}, \citenamefont {Rosi{\'n}ska}, \citenamefont {Ruggi}, \citenamefont {Saracco}, \citenamefont {Sassolas}, \citenamefont {Schimmel},
  \citenamefont {Sentenac}, \citenamefont {Sequino}, \citenamefont {Shah}, \citenamefont {Siellez}, \citenamefont {Straniero}, \citenamefont {Swinkels}, \citenamefont {Tacca}, \citenamefont {Tonelli}, \citenamefont {Travasso}, \citenamefont {Turconi}, \citenamefont {Vajente}, \citenamefont {van Bakel}, \citenamefont {van Beuzekom}, \citenamefont {van~den Brand}, \citenamefont {Broeck}, \citenamefont {van~der Sluys}, \citenamefont {van Heijningen}, \citenamefont {Vas{\'u}th}, \citenamefont {Vedovato}, \citenamefont {Veitch}, \citenamefont {Verkindt}, \citenamefont {Vetrano}, \citenamefont {Vicer{\'e}}, \citenamefont {Vinet}, \citenamefont {Visser}, \citenamefont {Vocca}, \citenamefont {Ward}, \citenamefont {Was}, \citenamefont {Wei}, \citenamefont {Yvert}, \citenamefont {{{\.z}ny}},\ and\ \citenamefont {Zendri}}]{acernese2014AdvancedVirgoSecondgeneration}%
  \BibitemOpen
  \bibfield  {author} {\bibinfo {author} {\bibfnamefont {F.}~\bibnamefont {Acernese}}, \bibinfo {author} {\bibfnamefont {M.}~\bibnamefont {Agathos}}, \bibinfo {author} {\bibfnamefont {K.}~\bibnamefont {Agatsuma}}, \bibinfo {author} {\bibfnamefont {D.}~\bibnamefont {Aisa}}, \bibinfo {author} {\bibfnamefont {N.}~\bibnamefont {Allemandou}}, \bibinfo {author} {\bibfnamefont {A.}~\bibnamefont {Allocca}}, \bibinfo {author} {\bibfnamefont {J.}~\bibnamefont {Amarni}}, \bibinfo {author} {\bibfnamefont {P.}~\bibnamefont {Astone}}, \bibinfo {author} {\bibfnamefont {G.}~\bibnamefont {Balestri}}, \bibinfo {author} {\bibfnamefont {G.}~\bibnamefont {Ballardin}}, \bibinfo {author} {\bibfnamefont {F.}~\bibnamefont {Barone}}, \bibinfo {author} {\bibfnamefont {J.-P.}\ \bibnamefont {Baronick}}, \bibinfo {author} {\bibfnamefont {M.}~\bibnamefont {Barsuglia}}, \bibinfo {author} {\bibfnamefont {A.}~\bibnamefont {Basti}}, \bibinfo {author} {\bibfnamefont {F.}~\bibnamefont {Basti}}, \bibinfo {author} {\bibfnamefont {T.~S.}\
  \bibnamefont {Bauer}}, \bibinfo {author} {\bibfnamefont {V.}~\bibnamefont {Bavigadda}}, \bibinfo {author} {\bibfnamefont {M.}~\bibnamefont {Bejger}}, \bibinfo {author} {\bibfnamefont {M.~G.}\ \bibnamefont {Beker}}, \bibinfo {author} {\bibfnamefont {C.}~\bibnamefont {Belczynski}}, \bibinfo {author} {\bibfnamefont {D.}~\bibnamefont {Bersanetti}}, \bibinfo {author} {\bibfnamefont {A.}~\bibnamefont {Bertolini}}, \bibinfo {author} {\bibfnamefont {M.}~\bibnamefont {Bitossi}}, \bibinfo {author} {\bibfnamefont {M.~A.}\ \bibnamefont {Bizouard}}, \bibinfo {author} {\bibfnamefont {S.}~\bibnamefont {Bloemen}}, \bibinfo {author} {\bibfnamefont {M.}~\bibnamefont {Blom}}, \bibinfo {author} {\bibfnamefont {M.}~\bibnamefont {Boer}}, \bibinfo {author} {\bibfnamefont {G.}~\bibnamefont {Bogaert}}, \bibinfo {author} {\bibfnamefont {D.}~\bibnamefont {Bondi}}, \bibinfo {author} {\bibfnamefont {F.}~\bibnamefont {Bondu}}, \bibinfo {author} {\bibfnamefont {L.}~\bibnamefont {Bonelli}}, \bibinfo {author} {\bibfnamefont
  {R.}~\bibnamefont {Bonnand}}, \bibinfo {author} {\bibfnamefont {V.}~\bibnamefont {Boschi}}, \bibinfo {author} {\bibfnamefont {L.}~\bibnamefont {Bosi}}, \bibinfo {author} {\bibfnamefont {T.}~\bibnamefont {Bouedo}}, \bibinfo {author} {\bibfnamefont {C.}~\bibnamefont {Bradaschia}}, \bibinfo {author} {\bibfnamefont {M.}~\bibnamefont {Branchesi}}, \bibinfo {author} {\bibfnamefont {T.}~\bibnamefont {Briant}}, \bibinfo {author} {\bibfnamefont {A.}~\bibnamefont {Brillet}}, \bibinfo {author} {\bibfnamefont {V.}~\bibnamefont {Brisson}}, \bibinfo {author} {\bibfnamefont {T.}~\bibnamefont {Bulik}}, \bibinfo {author} {\bibfnamefont {H.~J.}\ \bibnamefont {Bulten}}, \bibinfo {author} {\bibfnamefont {D.}~\bibnamefont {Buskulic}}, \bibinfo {author} {\bibfnamefont {C.}~\bibnamefont {Buy}}, \bibinfo {author} {\bibfnamefont {G.}~\bibnamefont {Cagnoli}}, \bibinfo {author} {\bibfnamefont {E.}~\bibnamefont {Calloni}}, \bibinfo {author} {\bibfnamefont {C.}~\bibnamefont {Campeggi}}, \bibinfo {author} {\bibfnamefont
  {B.}~\bibnamefont {Canuel}}, \bibinfo {author} {\bibfnamefont {F.}~\bibnamefont {Carbognani}}, \bibinfo {author} {\bibfnamefont {F.}~\bibnamefont {Cavalier}}, \bibinfo {author} {\bibfnamefont {R.}~\bibnamefont {Cavalieri}}, \bibinfo {author} {\bibfnamefont {G.}~\bibnamefont {Cella}}, \bibinfo {author} {\bibfnamefont {E.}~\bibnamefont {Cesarini}}, \bibinfo {author} {\bibfnamefont {E.~C.}\ \bibnamefont {Mottin}}, \bibinfo {author} {\bibfnamefont {A.}~\bibnamefont {Chincarini}}, \bibinfo {author} {\bibfnamefont {A.}~\bibnamefont {Chiummo}}, \bibinfo {author} {\bibfnamefont {S.}~\bibnamefont {Chua}}, \bibinfo {author} {\bibfnamefont {F.}~\bibnamefont {Cleva}}, \bibinfo {author} {\bibfnamefont {E.}~\bibnamefont {Coccia}}, \bibinfo {author} {\bibfnamefont {P.-F.}\ \bibnamefont {Cohadon}}, \bibinfo {author} {\bibfnamefont {A.}~\bibnamefont {Colla}}, \bibinfo {author} {\bibfnamefont {M.}~\bibnamefont {Colombini}}, \bibinfo {author} {\bibfnamefont {A.}~\bibnamefont {Conte}}, \bibinfo {author} {\bibfnamefont {J.-P.}\
  \bibnamefont {Coulon}}, \bibinfo {author} {\bibfnamefont {E.}~\bibnamefont {Cuoco}}, \bibinfo {author} {\bibfnamefont {A.}~\bibnamefont {Dalmaz}}, \bibinfo {author} {\bibfnamefont {S.}~\bibnamefont {D'Antonio}}, \bibinfo {author} {\bibfnamefont {V.}~\bibnamefont {Dattilo}}, \bibinfo {author} {\bibfnamefont {M.}~\bibnamefont {Davier}}, \bibinfo {author} {\bibfnamefont {R.}~\bibnamefont {Day}}, \bibinfo {author} {\bibfnamefont {G.}~\bibnamefont {Debreczeni}}, \bibinfo {author} {\bibfnamefont {J.}~\bibnamefont {Degallaix}}, \bibinfo {author} {\bibfnamefont {S.}~\bibnamefont {Del{\'e}glise}}, \bibinfo {author} {\bibfnamefont {W.~D.}\ \bibnamefont {Pozzo}}, \bibinfo {author} {\bibfnamefont {H.}~\bibnamefont {Dereli}}, \bibinfo {author} {\bibfnamefont {R.~D.}\ \bibnamefont {Rosa}}, \bibinfo {author} {\bibfnamefont {L.~D.}\ \bibnamefont {Fiore}}, \bibinfo {author} {\bibfnamefont {A.~D.}\ \bibnamefont {Lieto}}, \bibinfo {author} {\bibfnamefont {A.~D.}\ \bibnamefont {Virgilio}}, \bibinfo {author} {\bibfnamefont
  {M.}~\bibnamefont {Doets}}, \bibinfo {author} {\bibfnamefont {V.}~\bibnamefont {Dolique}}, \bibinfo {author} {\bibfnamefont {M.}~\bibnamefont {Drago}}, \bibinfo {author} {\bibfnamefont {M.}~\bibnamefont {Ducrot}}, \bibinfo {author} {\bibfnamefont {G.}~\bibnamefont {Endr{\H o}czi}}, \bibinfo {author} {\bibfnamefont {V.}~\bibnamefont {Fafone}}, \bibinfo {author} {\bibfnamefont {S.}~\bibnamefont {Farinon}}, \bibinfo {author} {\bibfnamefont {I.}~\bibnamefont {Ferrante}}, \bibinfo {author} {\bibfnamefont {F.}~\bibnamefont {Ferrini}}, \bibinfo {author} {\bibfnamefont {F.}~\bibnamefont {Fidecaro}}, \bibinfo {author} {\bibfnamefont {I.}~\bibnamefont {Fiori}}, \bibinfo {author} {\bibfnamefont {R.}~\bibnamefont {Flaminio}}, \bibinfo {author} {\bibfnamefont {J.-D.}\ \bibnamefont {Fournier}}, \bibinfo {author} {\bibfnamefont {S.}~\bibnamefont {Franco}}, \bibinfo {author} {\bibfnamefont {S.}~\bibnamefont {Frasca}}, \bibinfo {author} {\bibfnamefont {F.}~\bibnamefont {Frasconi}}, \bibinfo {author} {\bibfnamefont
  {L.}~\bibnamefont {Gammaitoni}}, \bibinfo {author} {\bibfnamefont {F.}~\bibnamefont {Garufi}}, \bibinfo {author} {\bibfnamefont {M.}~\bibnamefont {Gaspard}}, \bibinfo {author} {\bibfnamefont {A.}~\bibnamefont {Gatto}}, \bibinfo {author} {\bibfnamefont {G.}~\bibnamefont {Gemme}}, \bibinfo {author} {\bibfnamefont {B.}~\bibnamefont {Gendre}}, \bibinfo {author} {\bibfnamefont {E.}~\bibnamefont {Genin}}, \bibinfo {author} {\bibfnamefont {A.}~\bibnamefont {Gennai}}, \bibinfo {author} {\bibfnamefont {S.}~\bibnamefont {Ghosh}}, \bibinfo {author} {\bibfnamefont {L.}~\bibnamefont {Giacobone}}, \bibinfo {author} {\bibfnamefont {A.}~\bibnamefont {Giazotto}}, \bibinfo {author} {\bibfnamefont {R.}~\bibnamefont {Gouaty}}, \bibinfo {author} {\bibfnamefont {M.}~\bibnamefont {Granata}}, \bibinfo {author} {\bibfnamefont {G.}~\bibnamefont {Greco}}, \bibinfo {author} {\bibfnamefont {P.}~\bibnamefont {Groot}}, \bibinfo {author} {\bibfnamefont {G.~M.}\ \bibnamefont {Guidi}}, \bibinfo {author} {\bibfnamefont {J.}~\bibnamefont
  {Harms}}, \bibinfo {author} {\bibfnamefont {A.}~\bibnamefont {Heidmann}}, \bibinfo {author} {\bibfnamefont {H.}~\bibnamefont {Heitmann}}, \bibinfo {author} {\bibfnamefont {P.}~\bibnamefont {Hello}}, \bibinfo {author} {\bibfnamefont {G.}~\bibnamefont {Hemming}}, \bibinfo {author} {\bibfnamefont {E.}~\bibnamefont {Hennes}}, \bibinfo {author} {\bibfnamefont {D.}~\bibnamefont {Hofman}}, \bibinfo {author} {\bibfnamefont {P.}~\bibnamefont {Jaranowski}}, \bibinfo {author} {\bibfnamefont {R.~J.~G.}\ \bibnamefont {Jonker}}, \bibinfo {author} {\bibfnamefont {M.}~\bibnamefont {Kasprzack}}, \bibinfo {author} {\bibfnamefont {F.}~\bibnamefont {K{\'e}f{\'e}lian}}, \bibinfo {author} {\bibfnamefont {I.}~\bibnamefont {Kowalska}}, \bibinfo {author} {\bibfnamefont {M.}~\bibnamefont {Kraan}}, \bibinfo {author} {\bibfnamefont {A.}~\bibnamefont {Kr{\'o}lak}}, \bibinfo {author} {\bibfnamefont {A.}~\bibnamefont {Kutynia}}, \bibinfo {author} {\bibfnamefont {C.}~\bibnamefont {Lazzaro}}, \bibinfo {author} {\bibfnamefont
  {M.}~\bibnamefont {Leonardi}}, \bibinfo {author} {\bibfnamefont {N.}~\bibnamefont {Leroy}}, \bibinfo {author} {\bibfnamefont {N.}~\bibnamefont {Letendre}}, \bibinfo {author} {\bibfnamefont {T.~G.~F.}\ \bibnamefont {Li}}, \bibinfo {author} {\bibfnamefont {B.}~\bibnamefont {Lieunard}}, \bibinfo {author} {\bibfnamefont {M.}~\bibnamefont {Lorenzini}}, \bibinfo {author} {\bibfnamefont {V.}~\bibnamefont {Loriette}}, \bibinfo {author} {\bibfnamefont {G.}~\bibnamefont {Losurdo}}, \bibinfo {author} {\bibfnamefont {C.}~\bibnamefont {Magazz{\`u}}}, \bibinfo {author} {\bibfnamefont {E.}~\bibnamefont {Majorana}}, \bibinfo {author} {\bibfnamefont {I.}~\bibnamefont {Maksimovic}}, \bibinfo {author} {\bibfnamefont {V.}~\bibnamefont {Malvezzi}}, \bibinfo {author} {\bibfnamefont {N.}~\bibnamefont {Man}}, \bibinfo {author} {\bibfnamefont {V.}~\bibnamefont {Mangano}}, \bibinfo {author} {\bibfnamefont {M.}~\bibnamefont {Mantovani}}, \bibinfo {author} {\bibfnamefont {F.}~\bibnamefont {Marchesoni}}, \bibinfo {author}
  {\bibfnamefont {F.}~\bibnamefont {Marion}}, \bibinfo {author} {\bibfnamefont {J.}~\bibnamefont {Marque}}, \bibinfo {author} {\bibfnamefont {F.}~\bibnamefont {Martelli}}, \bibinfo {author} {\bibfnamefont {L.}~\bibnamefont {Martellini}}, \bibinfo {author} {\bibfnamefont {A.}~\bibnamefont {Masserot}}, \bibinfo {author} {\bibfnamefont {D.}~\bibnamefont {Meacher}}, \bibinfo {author} {\bibfnamefont {J.}~\bibnamefont {Meidam}}, \bibinfo {author} {\bibfnamefont {F.}~\bibnamefont {Mezzani}}, \bibinfo {author} {\bibfnamefont {C.}~\bibnamefont {Michel}}, \bibinfo {author} {\bibfnamefont {L.}~\bibnamefont {Milano}}, \bibinfo {author} {\bibfnamefont {Y.}~\bibnamefont {Minenkov}}, \bibinfo {author} {\bibfnamefont {A.}~\bibnamefont {Moggi}}, \bibinfo {author} {\bibfnamefont {M.}~\bibnamefont {Mohan}}, \bibinfo {author} {\bibfnamefont {M.}~\bibnamefont {Montani}}, \bibinfo {author} {\bibfnamefont {N.}~\bibnamefont {Morgado}}, \bibinfo {author} {\bibfnamefont {B.}~\bibnamefont {Mours}}, \bibinfo {author} {\bibfnamefont
  {F.}~\bibnamefont {Mul}}, \bibinfo {author} {\bibfnamefont {M.~F.}\ \bibnamefont {Nagy}}, \bibinfo {author} {\bibfnamefont {I.}~\bibnamefont {Nardecchia}}, \bibinfo {author} {\bibfnamefont {L.}~\bibnamefont {Naticchioni}}, \bibinfo {author} {\bibfnamefont {G.}~\bibnamefont {Nelemans}}, \bibinfo {author} {\bibfnamefont {I.}~\bibnamefont {Neri}}, \bibinfo {author} {\bibfnamefont {M.}~\bibnamefont {Neri}}, \bibinfo {author} {\bibfnamefont {F.}~\bibnamefont {Nocera}}, \bibinfo {author} {\bibfnamefont {E.}~\bibnamefont {Pacaud}}, \bibinfo {author} {\bibfnamefont {C.}~\bibnamefont {Palomba}}, \bibinfo {author} {\bibfnamefont {F.}~\bibnamefont {Paoletti}}, \bibinfo {author} {\bibfnamefont {A.}~\bibnamefont {Paoli}}, \bibinfo {author} {\bibfnamefont {A.}~\bibnamefont {Pasqualetti}}, \bibinfo {author} {\bibfnamefont {R.}~\bibnamefont {Passaquieti}}, \bibinfo {author} {\bibfnamefont {D.}~\bibnamefont {Passuello}}, \bibinfo {author} {\bibfnamefont {M.}~\bibnamefont {Perciballi}}, \bibinfo {author} {\bibfnamefont
  {S.}~\bibnamefont {Petit}}, \bibinfo {author} {\bibfnamefont {M.}~\bibnamefont {Pichot}}, \bibinfo {author} {\bibfnamefont {F.}~\bibnamefont {Piergiovanni}}, \bibinfo {author} {\bibfnamefont {G.}~\bibnamefont {Pillant}}, \bibinfo {author} {\bibfnamefont {A.}~\bibnamefont {Piluso}}, \bibinfo {author} {\bibfnamefont {L.}~\bibnamefont {Pinard}}, \bibinfo {author} {\bibfnamefont {R.}~\bibnamefont {Poggiani}}, \bibinfo {author} {\bibfnamefont {M.}~\bibnamefont {Prijatelj}}, \bibinfo {author} {\bibfnamefont {G.~A.}\ \bibnamefont {Prodi}}, \bibinfo {author} {\bibfnamefont {M.}~\bibnamefont {Punturo}}, \bibinfo {author} {\bibfnamefont {P.}~\bibnamefont {Puppo}}, \bibinfo {author} {\bibfnamefont {D.~S.}\ \bibnamefont {Rabeling}}, \bibinfo {author} {\bibfnamefont {I.}~\bibnamefont {R{\'a}cz}}, \bibinfo {author} {\bibfnamefont {P.}~\bibnamefont {Rapagnani}}, \bibinfo {author} {\bibfnamefont {M.}~\bibnamefont {Razzano}}, \bibinfo {author} {\bibfnamefont {V.}~\bibnamefont {Re}}, \bibinfo {author} {\bibfnamefont
  {T.}~\bibnamefont {Regimbau}}, \bibinfo {author} {\bibfnamefont {F.}~\bibnamefont {Ricci}}, \bibinfo {author} {\bibfnamefont {F.}~\bibnamefont {Robinet}}, \bibinfo {author} {\bibfnamefont {A.}~\bibnamefont {Rocchi}}, \bibinfo {author} {\bibfnamefont {L.}~\bibnamefont {Rolland}}, \bibinfo {author} {\bibfnamefont {R.}~\bibnamefont {Romano}}, \bibinfo {author} {\bibfnamefont {D.}~\bibnamefont {Rosi{\'n}ska}}, \bibinfo {author} {\bibfnamefont {P.}~\bibnamefont {Ruggi}}, \bibinfo {author} {\bibfnamefont {E.}~\bibnamefont {Saracco}}, \bibinfo {author} {\bibfnamefont {B.}~\bibnamefont {Sassolas}}, \bibinfo {author} {\bibfnamefont {F.}~\bibnamefont {Schimmel}}, \bibinfo {author} {\bibfnamefont {D.}~\bibnamefont {Sentenac}}, \bibinfo {author} {\bibfnamefont {V.}~\bibnamefont {Sequino}}, \bibinfo {author} {\bibfnamefont {S.}~\bibnamefont {Shah}}, \bibinfo {author} {\bibfnamefont {K.}~\bibnamefont {Siellez}}, \bibinfo {author} {\bibfnamefont {N.}~\bibnamefont {Straniero}}, \bibinfo {author} {\bibfnamefont
  {B.}~\bibnamefont {Swinkels}}, \bibinfo {author} {\bibfnamefont {M.}~\bibnamefont {Tacca}}, \bibinfo {author} {\bibfnamefont {M.}~\bibnamefont {Tonelli}}, \bibinfo {author} {\bibfnamefont {F.}~\bibnamefont {Travasso}}, \bibinfo {author} {\bibfnamefont {M.}~\bibnamefont {Turconi}}, \bibinfo {author} {\bibfnamefont {G.}~\bibnamefont {Vajente}}, \bibinfo {author} {\bibfnamefont {N.}~\bibnamefont {van Bakel}}, \bibinfo {author} {\bibfnamefont {M.}~\bibnamefont {van Beuzekom}}, \bibinfo {author} {\bibfnamefont {J.~F.~J.}\ \bibnamefont {van~den Brand}}, \bibinfo {author} {\bibfnamefont {C.~V.~D.}\ \bibnamefont {Broeck}}, \bibinfo {author} {\bibfnamefont {M.~V.}\ \bibnamefont {van~der Sluys}}, \bibinfo {author} {\bibfnamefont {J.}~\bibnamefont {van Heijningen}}, \bibinfo {author} {\bibfnamefont {M.}~\bibnamefont {Vas{\'u}th}}, \bibinfo {author} {\bibfnamefont {G.}~\bibnamefont {Vedovato}}, \bibinfo {author} {\bibfnamefont {J.}~\bibnamefont {Veitch}}, \bibinfo {author} {\bibfnamefont {D.}~\bibnamefont {Verkindt}},
  \bibinfo {author} {\bibfnamefont {F.}~\bibnamefont {Vetrano}}, \bibinfo {author} {\bibfnamefont {A.}~\bibnamefont {Vicer{\'e}}}, \bibinfo {author} {\bibfnamefont {J.-Y.}\ \bibnamefont {Vinet}}, \bibinfo {author} {\bibfnamefont {G.}~\bibnamefont {Visser}}, \bibinfo {author} {\bibfnamefont {H.}~\bibnamefont {Vocca}}, \bibinfo {author} {\bibfnamefont {R.}~\bibnamefont {Ward}}, \bibinfo {author} {\bibfnamefont {M.}~\bibnamefont {Was}}, \bibinfo {author} {\bibfnamefont {L.-W.}\ \bibnamefont {Wei}}, \bibinfo {author} {\bibfnamefont {M.}~\bibnamefont {Yvert}}, \bibinfo {author} {\bibfnamefont {A.~Z.}\ \bibnamefont {{{\.z}ny}}},\ and\ \bibinfo {author} {\bibfnamefont {J.-P.}\ \bibnamefont {Zendri}},\ }\href {https://doi.org/10.1088/0264-9381/32/2/024001} {\bibfield  {journal} {\bibinfo  {journal} {Class. Quantum Grav.}\ }\textbf {\bibinfo {volume} {32}},\ \bibinfo {pages} {024001} (\bibinfo {year} {2014})}\BibitemShut {NoStop}%
\bibitem [{\citenamefont {Akutsu}\ \emph {et~al.}(2021)\citenamefont {Akutsu}, \citenamefont {Ando}, \citenamefont {Arai}, \citenamefont {Arai}, \citenamefont {Araki}, \citenamefont {Araya}, \citenamefont {Aritomi}, \citenamefont {Aso}, \citenamefont {Bae}, \citenamefont {Bae}, \citenamefont {Baiotti}, \citenamefont {Bajpai}, \citenamefont {Barton}, \citenamefont {Cannon}, \citenamefont {Capocasa}, \citenamefont {Chan}, \citenamefont {Chen}, \citenamefont {Chen}, \citenamefont {Chen}, \citenamefont {Chu}, \citenamefont {Chu}, \citenamefont {Eguchi}, \citenamefont {Enomoto}, \citenamefont {Flaminio}, \citenamefont {Fujii}, \citenamefont {Fukunaga}, \citenamefont {Fukushima}, \citenamefont {Ge}, \citenamefont {Hagiwara}, \citenamefont {Haino}, \citenamefont {Hasegawa}, \citenamefont {Hayakawa}, \citenamefont {Hayama}, \citenamefont {Himemoto}, \citenamefont {Hiranuma}, \citenamefont {Hirata}, \citenamefont {Hirose}, \citenamefont {Hong}, \citenamefont {Hsieh}, \citenamefont {Huang}, \citenamefont {Huang},
  \citenamefont {Huang}, \citenamefont {Ikenoue}, \citenamefont {Imam}, \citenamefont {Inayoshi}, \citenamefont {Inoue}, \citenamefont {Ioka}, \citenamefont {Itoh}, \citenamefont {Izumi}, \citenamefont {Jung}, \citenamefont {Jung}, \citenamefont {Kajita}, \citenamefont {Kamiizumi}, \citenamefont {Kanda}, \citenamefont {Kang}, \citenamefont {Kawaguchi}, \citenamefont {Kawai}, \citenamefont {Kawasaki}, \citenamefont {Kim}, \citenamefont {Kim}, \citenamefont {Kim}, \citenamefont {Kim}, \citenamefont {Kimura}, \citenamefont {Kita}, \citenamefont {Kitazawa}, \citenamefont {Kojima}, \citenamefont {Kokeyama}, \citenamefont {Komori}, \citenamefont {Kong}, \citenamefont {Kotake}, \citenamefont {Kozakai}, \citenamefont {Kozu}, \citenamefont {Kumar}, \citenamefont {Kume}, \citenamefont {Kuo}, \citenamefont {Kuo}, \citenamefont {Kuroyanagi}, \citenamefont {Kusayanagi}, \citenamefont {Kwak}, \citenamefont {Lee}, \citenamefont {Lee}, \citenamefont {Lee}, \citenamefont {Leonardi}, \citenamefont {Lin}, \citenamefont {Lin},
  \citenamefont {Lin}, \citenamefont {Liu}, \citenamefont {Luo}, \citenamefont {Marchio}, \citenamefont {Michimura}, \citenamefont {Mio}, \citenamefont {Miyakawa}, \citenamefont {Miyamoto}, \citenamefont {Miyazaki}, \citenamefont {Miyo}, \citenamefont {Miyoki}, \citenamefont {Morisaki}, \citenamefont {Moriwaki}, \citenamefont {Nagano}, \citenamefont {Nagano}, \citenamefont {Nakamura}, \citenamefont {Nakano}, \citenamefont {Nakano}, \citenamefont {Nakashima}, \citenamefont {Narikawa}, \citenamefont {Negishi}, \citenamefont {Ni}, \citenamefont {Nishizawa}, \citenamefont {Obuchi}, \citenamefont {Ogaki}, \citenamefont {Oh}, \citenamefont {Oh}, \citenamefont {Ohashi}, \citenamefont {Ohishi}, \citenamefont {Ohkawa}, \citenamefont {Okutomi}, \citenamefont {Oohara}, \citenamefont {Ooi}, \citenamefont {Oshino}, \citenamefont {Pan}, \citenamefont {Pang}, \citenamefont {Park}, \citenamefont {Arellano}, \citenamefont {Pinto}, \citenamefont {Sago}, \citenamefont {Saito}, \citenamefont {Saito}, \citenamefont {Sakai},
  \citenamefont {Sakai}, \citenamefont {Sakuno}, \citenamefont {Sato}, \citenamefont {Sato}, \citenamefont {Sawada}, \citenamefont {Sekiguchi}, \citenamefont {Sekiguchi}, \citenamefont {Shibagaki}, \citenamefont {Shimizu}, \citenamefont {Shimoda}, \citenamefont {Shimode}, \citenamefont {Shinkai}, \citenamefont {Shishido}, \citenamefont {Shoda}, \citenamefont {Somiya}, \citenamefont {Son}, \citenamefont {Sotani}, \citenamefont {Sugimoto}, \citenamefont {Suzuki}, \citenamefont {Suzuki}, \citenamefont {Tagoshi}, \citenamefont {Takahashi}, \citenamefont {Takahashi}, \citenamefont {Takamori}, \citenamefont {Takano}, \citenamefont {Takeda}, \citenamefont {Takeda}, \citenamefont {Tanaka}, \citenamefont {Tanaka}, \citenamefont {Tanaka}, \citenamefont {Tanaka}, \citenamefont {Tanaka}, \citenamefont {Tanioka}, \citenamefont {Tapia San~Martin}, \citenamefont {Telada}, \citenamefont {Tomaru}, \citenamefont {Tomigami}, \citenamefont {Tomura}, \citenamefont {Travasso}, \citenamefont {Trozzo}, \citenamefont {Tsang},
  \citenamefont {Tsubono}, \citenamefont {Tsuchida}, \citenamefont {Tsuzuki}, \citenamefont {Tuyenbayev}, \citenamefont {Uchikata}, \citenamefont {Uchiyama}, \citenamefont {Ueda}, \citenamefont {Uehara}, \citenamefont {Ueno}, \citenamefont {Ueshima}, \citenamefont {Uraguchi}, \citenamefont {Ushiba}, \citenamefont {{van Putten}}, \citenamefont {Vocca}, \citenamefont {Wang}, \citenamefont {Wu}, \citenamefont {Wu}, \citenamefont {Wu}, \citenamefont {Xu}, \citenamefont {Yamada}, \citenamefont {Yamamoto}, \citenamefont {Yamamoto}, \citenamefont {Yamamoto}, \citenamefont {Yokogawa}, \citenamefont {Yokoyama}, \citenamefont {Yokozawa}, \citenamefont {Yoshioka}, \citenamefont {Yuzurihara}, \citenamefont {Zeidler}, \citenamefont {Zhao},\ and\ \citenamefont {Zhu}}]{akutsu2021OverviewKAGRADetector}%
  \BibitemOpen
  \bibfield  {author} {\bibinfo {author} {\bibfnamefont {T.}~\bibnamefont {Akutsu}}, \bibinfo {author} {\bibfnamefont {M.}~\bibnamefont {Ando}}, \bibinfo {author} {\bibfnamefont {K.}~\bibnamefont {Arai}}, \bibinfo {author} {\bibfnamefont {Y.}~\bibnamefont {Arai}}, \bibinfo {author} {\bibfnamefont {S.}~\bibnamefont {Araki}}, \bibinfo {author} {\bibfnamefont {A.}~\bibnamefont {Araya}}, \bibinfo {author} {\bibfnamefont {N.}~\bibnamefont {Aritomi}}, \bibinfo {author} {\bibfnamefont {Y.}~\bibnamefont {Aso}}, \bibinfo {author} {\bibfnamefont {S.}~\bibnamefont {Bae}}, \bibinfo {author} {\bibfnamefont {Y.}~\bibnamefont {Bae}}, \bibinfo {author} {\bibfnamefont {L.}~\bibnamefont {Baiotti}}, \bibinfo {author} {\bibfnamefont {R.}~\bibnamefont {Bajpai}}, \bibinfo {author} {\bibfnamefont {M.~A.}\ \bibnamefont {Barton}}, \bibinfo {author} {\bibfnamefont {K.}~\bibnamefont {Cannon}}, \bibinfo {author} {\bibfnamefont {E.}~\bibnamefont {Capocasa}}, \bibinfo {author} {\bibfnamefont {M.}~\bibnamefont {Chan}}, \bibinfo {author}
  {\bibfnamefont {C.}~\bibnamefont {Chen}}, \bibinfo {author} {\bibfnamefont {K.}~\bibnamefont {Chen}}, \bibinfo {author} {\bibfnamefont {Y.}~\bibnamefont {Chen}}, \bibinfo {author} {\bibfnamefont {H.}~\bibnamefont {Chu}}, \bibinfo {author} {\bibfnamefont {Y.~K.}\ \bibnamefont {Chu}}, \bibinfo {author} {\bibfnamefont {S.}~\bibnamefont {Eguchi}}, \bibinfo {author} {\bibfnamefont {Y.}~\bibnamefont {Enomoto}}, \bibinfo {author} {\bibfnamefont {R.}~\bibnamefont {Flaminio}}, \bibinfo {author} {\bibfnamefont {Y.}~\bibnamefont {Fujii}}, \bibinfo {author} {\bibfnamefont {M.}~\bibnamefont {Fukunaga}}, \bibinfo {author} {\bibfnamefont {M.}~\bibnamefont {Fukushima}}, \bibinfo {author} {\bibfnamefont {G.}~\bibnamefont {Ge}}, \bibinfo {author} {\bibfnamefont {A.}~\bibnamefont {Hagiwara}}, \bibinfo {author} {\bibfnamefont {S.}~\bibnamefont {Haino}}, \bibinfo {author} {\bibfnamefont {K.}~\bibnamefont {Hasegawa}}, \bibinfo {author} {\bibfnamefont {H.}~\bibnamefont {Hayakawa}}, \bibinfo {author} {\bibfnamefont
  {K.}~\bibnamefont {Hayama}}, \bibinfo {author} {\bibfnamefont {Y.}~\bibnamefont {Himemoto}}, \bibinfo {author} {\bibfnamefont {Y.}~\bibnamefont {Hiranuma}}, \bibinfo {author} {\bibfnamefont {N.}~\bibnamefont {Hirata}}, \bibinfo {author} {\bibfnamefont {E.}~\bibnamefont {Hirose}}, \bibinfo {author} {\bibfnamefont {Z.}~\bibnamefont {Hong}}, \bibinfo {author} {\bibfnamefont {B.~H.}\ \bibnamefont {Hsieh}}, \bibinfo {author} {\bibfnamefont {C.~Z.}\ \bibnamefont {Huang}}, \bibinfo {author} {\bibfnamefont {P.}~\bibnamefont {Huang}}, \bibinfo {author} {\bibfnamefont {Y.}~\bibnamefont {Huang}}, \bibinfo {author} {\bibfnamefont {B.}~\bibnamefont {Ikenoue}}, \bibinfo {author} {\bibfnamefont {S.}~\bibnamefont {Imam}}, \bibinfo {author} {\bibfnamefont {K.}~\bibnamefont {Inayoshi}}, \bibinfo {author} {\bibfnamefont {Y.}~\bibnamefont {Inoue}}, \bibinfo {author} {\bibfnamefont {K.}~\bibnamefont {Ioka}}, \bibinfo {author} {\bibfnamefont {Y.}~\bibnamefont {Itoh}}, \bibinfo {author} {\bibfnamefont {K.}~\bibnamefont {Izumi}},
  \bibinfo {author} {\bibfnamefont {K.}~\bibnamefont {Jung}}, \bibinfo {author} {\bibfnamefont {P.}~\bibnamefont {Jung}}, \bibinfo {author} {\bibfnamefont {T.}~\bibnamefont {Kajita}}, \bibinfo {author} {\bibfnamefont {M.}~\bibnamefont {Kamiizumi}}, \bibinfo {author} {\bibfnamefont {N.}~\bibnamefont {Kanda}}, \bibinfo {author} {\bibfnamefont {G.}~\bibnamefont {Kang}}, \bibinfo {author} {\bibfnamefont {K.}~\bibnamefont {Kawaguchi}}, \bibinfo {author} {\bibfnamefont {N.}~\bibnamefont {Kawai}}, \bibinfo {author} {\bibfnamefont {T.}~\bibnamefont {Kawasaki}}, \bibinfo {author} {\bibfnamefont {C.}~\bibnamefont {Kim}}, \bibinfo {author} {\bibfnamefont {J.~C.}\ \bibnamefont {Kim}}, \bibinfo {author} {\bibfnamefont {W.~S.}\ \bibnamefont {Kim}}, \bibinfo {author} {\bibfnamefont {Y.~M.}\ \bibnamefont {Kim}}, \bibinfo {author} {\bibfnamefont {N.}~\bibnamefont {Kimura}}, \bibinfo {author} {\bibfnamefont {N.}~\bibnamefont {Kita}}, \bibinfo {author} {\bibfnamefont {H.}~\bibnamefont {Kitazawa}}, \bibinfo {author}
  {\bibfnamefont {Y.}~\bibnamefont {Kojima}}, \bibinfo {author} {\bibfnamefont {K.}~\bibnamefont {Kokeyama}}, \bibinfo {author} {\bibfnamefont {K.}~\bibnamefont {Komori}}, \bibinfo {author} {\bibfnamefont {A.~K.~H.}\ \bibnamefont {Kong}}, \bibinfo {author} {\bibfnamefont {K.}~\bibnamefont {Kotake}}, \bibinfo {author} {\bibfnamefont {C.}~\bibnamefont {Kozakai}}, \bibinfo {author} {\bibfnamefont {R.}~\bibnamefont {Kozu}}, \bibinfo {author} {\bibfnamefont {R.}~\bibnamefont {Kumar}}, \bibinfo {author} {\bibfnamefont {J.}~\bibnamefont {Kume}}, \bibinfo {author} {\bibfnamefont {C.}~\bibnamefont {Kuo}}, \bibinfo {author} {\bibfnamefont {H.~S.}\ \bibnamefont {Kuo}}, \bibinfo {author} {\bibfnamefont {S.}~\bibnamefont {Kuroyanagi}}, \bibinfo {author} {\bibfnamefont {K.}~\bibnamefont {Kusayanagi}}, \bibinfo {author} {\bibfnamefont {K.}~\bibnamefont {Kwak}}, \bibinfo {author} {\bibfnamefont {H.~K.}\ \bibnamefont {Lee}}, \bibinfo {author} {\bibfnamefont {H.~W.}\ \bibnamefont {Lee}}, \bibinfo {author} {\bibfnamefont
  {R.}~\bibnamefont {Lee}}, \bibinfo {author} {\bibfnamefont {M.}~\bibnamefont {Leonardi}}, \bibinfo {author} {\bibfnamefont {L.~C.~C.}\ \bibnamefont {Lin}}, \bibinfo {author} {\bibfnamefont {C.~Y.}\ \bibnamefont {Lin}}, \bibinfo {author} {\bibfnamefont {F.~L.}\ \bibnamefont {Lin}}, \bibinfo {author} {\bibfnamefont {G.~C.}\ \bibnamefont {Liu}}, \bibinfo {author} {\bibfnamefont {L.~W.}\ \bibnamefont {Luo}}, \bibinfo {author} {\bibfnamefont {M.}~\bibnamefont {Marchio}}, \bibinfo {author} {\bibfnamefont {Y.}~\bibnamefont {Michimura}}, \bibinfo {author} {\bibfnamefont {N.}~\bibnamefont {Mio}}, \bibinfo {author} {\bibfnamefont {O.}~\bibnamefont {Miyakawa}}, \bibinfo {author} {\bibfnamefont {A.}~\bibnamefont {Miyamoto}}, \bibinfo {author} {\bibfnamefont {Y.}~\bibnamefont {Miyazaki}}, \bibinfo {author} {\bibfnamefont {K.}~\bibnamefont {Miyo}}, \bibinfo {author} {\bibfnamefont {S.}~\bibnamefont {Miyoki}}, \bibinfo {author} {\bibfnamefont {S.}~\bibnamefont {Morisaki}}, \bibinfo {author} {\bibfnamefont
  {Y.}~\bibnamefont {Moriwaki}}, \bibinfo {author} {\bibfnamefont {K.}~\bibnamefont {Nagano}}, \bibinfo {author} {\bibfnamefont {S.}~\bibnamefont {Nagano}}, \bibinfo {author} {\bibfnamefont {K.}~\bibnamefont {Nakamura}}, \bibinfo {author} {\bibfnamefont {H.}~\bibnamefont {Nakano}}, \bibinfo {author} {\bibfnamefont {M.}~\bibnamefont {Nakano}}, \bibinfo {author} {\bibfnamefont {R.}~\bibnamefont {Nakashima}}, \bibinfo {author} {\bibfnamefont {T.}~\bibnamefont {Narikawa}}, \bibinfo {author} {\bibfnamefont {R.}~\bibnamefont {Negishi}}, \bibinfo {author} {\bibfnamefont {W.~T.}\ \bibnamefont {Ni}}, \bibinfo {author} {\bibfnamefont {A.}~\bibnamefont {Nishizawa}}, \bibinfo {author} {\bibfnamefont {Y.}~\bibnamefont {Obuchi}}, \bibinfo {author} {\bibfnamefont {W.}~\bibnamefont {Ogaki}}, \bibinfo {author} {\bibfnamefont {J.~J.}\ \bibnamefont {Oh}}, \bibinfo {author} {\bibfnamefont {S.~H.}\ \bibnamefont {Oh}}, \bibinfo {author} {\bibfnamefont {M.}~\bibnamefont {Ohashi}}, \bibinfo {author} {\bibfnamefont {N.}~\bibnamefont
  {Ohishi}}, \bibinfo {author} {\bibfnamefont {M.}~\bibnamefont {Ohkawa}}, \bibinfo {author} {\bibfnamefont {K.}~\bibnamefont {Okutomi}}, \bibinfo {author} {\bibfnamefont {K.}~\bibnamefont {Oohara}}, \bibinfo {author} {\bibfnamefont {C.~P.}\ \bibnamefont {Ooi}}, \bibinfo {author} {\bibfnamefont {S.}~\bibnamefont {Oshino}}, \bibinfo {author} {\bibfnamefont {K.}~\bibnamefont {Pan}}, \bibinfo {author} {\bibfnamefont {H.}~\bibnamefont {Pang}}, \bibinfo {author} {\bibfnamefont {J.}~\bibnamefont {Park}}, \bibinfo {author} {\bibfnamefont {F.~E.~P.}\ \bibnamefont {Arellano}}, \bibinfo {author} {\bibfnamefont {I.}~\bibnamefont {Pinto}}, \bibinfo {author} {\bibfnamefont {N.}~\bibnamefont {Sago}}, \bibinfo {author} {\bibfnamefont {S.}~\bibnamefont {Saito}}, \bibinfo {author} {\bibfnamefont {Y.}~\bibnamefont {Saito}}, \bibinfo {author} {\bibfnamefont {K.}~\bibnamefont {Sakai}}, \bibinfo {author} {\bibfnamefont {Y.}~\bibnamefont {Sakai}}, \bibinfo {author} {\bibfnamefont {Y.}~\bibnamefont {Sakuno}}, \bibinfo {author}
  {\bibfnamefont {S.}~\bibnamefont {Sato}}, \bibinfo {author} {\bibfnamefont {T.}~\bibnamefont {Sato}}, \bibinfo {author} {\bibfnamefont {T.}~\bibnamefont {Sawada}}, \bibinfo {author} {\bibfnamefont {T.}~\bibnamefont {Sekiguchi}}, \bibinfo {author} {\bibfnamefont {Y.}~\bibnamefont {Sekiguchi}}, \bibinfo {author} {\bibfnamefont {S.}~\bibnamefont {Shibagaki}}, \bibinfo {author} {\bibfnamefont {R.}~\bibnamefont {Shimizu}}, \bibinfo {author} {\bibfnamefont {T.}~\bibnamefont {Shimoda}}, \bibinfo {author} {\bibfnamefont {K.}~\bibnamefont {Shimode}}, \bibinfo {author} {\bibfnamefont {H.}~\bibnamefont {Shinkai}}, \bibinfo {author} {\bibfnamefont {T.}~\bibnamefont {Shishido}}, \bibinfo {author} {\bibfnamefont {A.}~\bibnamefont {Shoda}}, \bibinfo {author} {\bibfnamefont {K.}~\bibnamefont {Somiya}}, \bibinfo {author} {\bibfnamefont {E.~J.}\ \bibnamefont {Son}}, \bibinfo {author} {\bibfnamefont {H.}~\bibnamefont {Sotani}}, \bibinfo {author} {\bibfnamefont {R.}~\bibnamefont {Sugimoto}}, \bibinfo {author} {\bibfnamefont
  {T.}~\bibnamefont {Suzuki}}, \bibinfo {author} {\bibfnamefont {T.}~\bibnamefont {Suzuki}}, \bibinfo {author} {\bibfnamefont {H.}~\bibnamefont {Tagoshi}}, \bibinfo {author} {\bibfnamefont {H.}~\bibnamefont {Takahashi}}, \bibinfo {author} {\bibfnamefont {R.}~\bibnamefont {Takahashi}}, \bibinfo {author} {\bibfnamefont {A.}~\bibnamefont {Takamori}}, \bibinfo {author} {\bibfnamefont {S.}~\bibnamefont {Takano}}, \bibinfo {author} {\bibfnamefont {H.}~\bibnamefont {Takeda}}, \bibinfo {author} {\bibfnamefont {M.}~\bibnamefont {Takeda}}, \bibinfo {author} {\bibfnamefont {H.}~\bibnamefont {Tanaka}}, \bibinfo {author} {\bibfnamefont {K.}~\bibnamefont {Tanaka}}, \bibinfo {author} {\bibfnamefont {K.}~\bibnamefont {Tanaka}}, \bibinfo {author} {\bibfnamefont {T.}~\bibnamefont {Tanaka}}, \bibinfo {author} {\bibfnamefont {T.}~\bibnamefont {Tanaka}}, \bibinfo {author} {\bibfnamefont {S.}~\bibnamefont {Tanioka}}, \bibinfo {author} {\bibfnamefont {E.~N.}\ \bibnamefont {Tapia San~Martin}}, \bibinfo {author} {\bibfnamefont
  {S.}~\bibnamefont {Telada}}, \bibinfo {author} {\bibfnamefont {T.}~\bibnamefont {Tomaru}}, \bibinfo {author} {\bibfnamefont {Y.}~\bibnamefont {Tomigami}}, \bibinfo {author} {\bibfnamefont {T.}~\bibnamefont {Tomura}}, \bibinfo {author} {\bibfnamefont {F.}~\bibnamefont {Travasso}}, \bibinfo {author} {\bibfnamefont {L.}~\bibnamefont {Trozzo}}, \bibinfo {author} {\bibfnamefont {T.}~\bibnamefont {Tsang}}, \bibinfo {author} {\bibfnamefont {K.}~\bibnamefont {Tsubono}}, \bibinfo {author} {\bibfnamefont {S.}~\bibnamefont {Tsuchida}}, \bibinfo {author} {\bibfnamefont {T.}~\bibnamefont {Tsuzuki}}, \bibinfo {author} {\bibfnamefont {D.}~\bibnamefont {Tuyenbayev}}, \bibinfo {author} {\bibfnamefont {N.}~\bibnamefont {Uchikata}}, \bibinfo {author} {\bibfnamefont {T.}~\bibnamefont {Uchiyama}}, \bibinfo {author} {\bibfnamefont {A.}~\bibnamefont {Ueda}}, \bibinfo {author} {\bibfnamefont {T.}~\bibnamefont {Uehara}}, \bibinfo {author} {\bibfnamefont {K.}~\bibnamefont {Ueno}}, \bibinfo {author} {\bibfnamefont {G.}~\bibnamefont
  {Ueshima}}, \bibinfo {author} {\bibfnamefont {F.}~\bibnamefont {Uraguchi}}, \bibinfo {author} {\bibfnamefont {T.}~\bibnamefont {Ushiba}}, \bibinfo {author} {\bibfnamefont {M.~H. P.~M.}\ \bibnamefont {{van Putten}}}, \bibinfo {author} {\bibfnamefont {H.}~\bibnamefont {Vocca}}, \bibinfo {author} {\bibfnamefont {J.}~\bibnamefont {Wang}}, \bibinfo {author} {\bibfnamefont {C.}~\bibnamefont {Wu}}, \bibinfo {author} {\bibfnamefont {H.}~\bibnamefont {Wu}}, \bibinfo {author} {\bibfnamefont {S.}~\bibnamefont {Wu}}, \bibinfo {author} {\bibfnamefont {W.-R.}\ \bibnamefont {Xu}}, \bibinfo {author} {\bibfnamefont {T.}~\bibnamefont {Yamada}}, \bibinfo {author} {\bibfnamefont {K.}~\bibnamefont {Yamamoto}}, \bibinfo {author} {\bibfnamefont {K.}~\bibnamefont {Yamamoto}}, \bibinfo {author} {\bibfnamefont {T.}~\bibnamefont {Yamamoto}}, \bibinfo {author} {\bibfnamefont {K.}~\bibnamefont {Yokogawa}}, \bibinfo {author} {\bibfnamefont {J.}~\bibnamefont {Yokoyama}}, \bibinfo {author} {\bibfnamefont {T.}~\bibnamefont {Yokozawa}},
  \bibinfo {author} {\bibfnamefont {T.}~\bibnamefont {Yoshioka}}, \bibinfo {author} {\bibfnamefont {H.}~\bibnamefont {Yuzurihara}}, \bibinfo {author} {\bibfnamefont {S.}~\bibnamefont {Zeidler}}, \bibinfo {author} {\bibfnamefont {Y.}~\bibnamefont {Zhao}},\ and\ \bibinfo {author} {\bibfnamefont {Z.~H.}\ \bibnamefont {Zhu}},\ }\href {https://doi.org/10.1093/ptep/ptaa125} {\bibfield  {journal} {\bibinfo  {journal} {Progress of Theoretical and Experimental Physics}\ }\textbf {\bibinfo {volume} {2021}},\ \bibinfo {pages} {05A101} (\bibinfo {year} {2021})}\BibitemShut {NoStop}%
\bibitem [{\citenamefont {{The KAGRA Collaboration}}\ \emph {et~al.}(2013)\citenamefont {{The KAGRA Collaboration}}, \citenamefont {Aso}, \citenamefont {Michimura}, \citenamefont {Somiya}, \citenamefont {Ando}, \citenamefont {Miyakawa}, \citenamefont {Sekiguchi}, \citenamefont {Tatsumi},\ and\ \citenamefont {Yamamoto}}]{thekagracollaboration2013InterferometerDesignKAGRA}%
  \BibitemOpen
  \bibfield  {author} {\bibinfo {author} {\bibnamefont {{The KAGRA Collaboration}}}, \bibinfo {author} {\bibfnamefont {Y.}~\bibnamefont {Aso}}, \bibinfo {author} {\bibfnamefont {Y.}~\bibnamefont {Michimura}}, \bibinfo {author} {\bibfnamefont {K.}~\bibnamefont {Somiya}}, \bibinfo {author} {\bibfnamefont {M.}~\bibnamefont {Ando}}, \bibinfo {author} {\bibfnamefont {O.}~\bibnamefont {Miyakawa}}, \bibinfo {author} {\bibfnamefont {T.}~\bibnamefont {Sekiguchi}}, \bibinfo {author} {\bibfnamefont {D.}~\bibnamefont {Tatsumi}},\ and\ \bibinfo {author} {\bibfnamefont {H.}~\bibnamefont {Yamamoto}},\ }\href {https://doi.org/10.1103/PhysRevD.88.043007} {\bibfield  {journal} {\bibinfo  {journal} {Phys. Rev. D}\ }\textbf {\bibinfo {volume} {88}},\ \bibinfo {pages} {043007} (\bibinfo {year} {2013})}\BibitemShut {NoStop}%
\bibitem [{\citenamefont {Junker}\ \emph {et~al.}(2022)\citenamefont {Junker}, \citenamefont {Wilken}, \citenamefont {Johny}, \citenamefont {Steinmeyer},\ and\ \citenamefont {Heurs}}]{junker2022FrequencyDependentSqueezingDetuned}%
  \BibitemOpen
  \bibfield  {author} {\bibinfo {author} {\bibfnamefont {J.}~\bibnamefont {Junker}}, \bibinfo {author} {\bibfnamefont {D.}~\bibnamefont {Wilken}}, \bibinfo {author} {\bibfnamefont {N.}~\bibnamefont {Johny}}, \bibinfo {author} {\bibfnamefont {D.}~\bibnamefont {Steinmeyer}},\ and\ \bibinfo {author} {\bibfnamefont {M.}~\bibnamefont {Heurs}},\ }\href {https://doi.org/10.1103/PhysRevLett.129.033602} {\bibfield  {journal} {\bibinfo  {journal} {Phys. Rev. Lett.}\ }\textbf {\bibinfo {volume} {129}},\ \bibinfo {pages} {033602} (\bibinfo {year} {2022})}\BibitemShut {NoStop}%
\end{thebibliography}

%

\end{document}